\documentclass{tCPH2e}

\usepackage{epstopdf}
\usepackage{subfigure}
\usepackage{siunitx}
\usepackage{hyperref}
\usepackage[utf8x]{inputenc}
\usepackage[english]{babel}

\usepackage{amssymb,amsmath}
\usepackage{bm}
\usepackage{color}
\usepackage{wrapfig}

\theoremstyle{plain}

\theoremstyle{definition}

\theoremstyle{remark}

\newcommand{\D}{{\mathrm{d}}}

\newcommand{\ff}{\mbox{\boldmath$f$}}

\begin{document}


\articletype{Review article}

\title{Beyond Navier--Stokes equations: Capillarity of ideal gas}

\author{
\name{A.N. Gorban\textsuperscript{a}$^{\ast}$\thanks{$^\ast$Corresponding author. Email: ag153@le.ac.uk}
and I.V. Karlin\textsuperscript{b}$^\ddag$\thanks{$^\ddag$Email: karlin@lav.mavt.ethz.ch}}
\affil{\textsuperscript{a}Department of Mathematics, University of Leicester, LE1 7RH Leicester, UK;
\textsuperscript{b}Department of Mechanical and Process Engineering,
ETH Z\"urich,
CH-8092, Z\"urich, Switzerland}
\received{2016}
}

\maketitle

{\bf Published in Contemporary Physics,  58(1) (2017), 70--90}

\vspace{5mm}

\begin{abstract}
The system of Navier--Stokes--Fourier  equations is one of the most celebrated systems of equations in modern science. It describes
dynamics of fluids in the limit when gradients of density, velocity and temperature are sufficiently small,  and  loses its applicability when the flux becomes so non-equilibrium that the changes of velocity, density or temperature on the length compatible with the mean free path are non-negligible. The question is: how to model such fluxes? This problem is still open. (Despite the fact that the first `final equations of motion' modified for analysis of thermal creep in rarefied gas were proposed by Maxwell in 1879.) There are, at least, three possible answers: (i) use molecular dynamics with individual particles, (ii) use kinetic equations, like Boltzmann's equation, or (iii) find a new system of equations for description of fluid dynamics with better accounting of non-equilibrium effects. These three approaches work at different scales. We explore the third possibility using the recent findings of capillarity of internal layers in ideal gases and of saturation effect in dissipation (there is a limiting attenuation rate for very short waves in ideal gas and it cannot increase infinitely). One candidate equation is discussed in more detail, the Korteweg system proposed in 1901.  The main ideas and approaches are illustrated by a kinetic system for which the problem of reduction of  kinetics to fluid dynamics is analytically solvable.
\end{abstract}


\section{Introduction}
\subsection{Timeline (the history of equations)}
Historical timelines are seldom  comprehensive and are often subject to debates but they do provide a background to examine the present. Older chronology seems to be more reliable and less disputable (it may happen because some details went forgotten). For understanding the situation `beyond Navier--Stokes' the following milestones from the distant past  are crucial:
\begin{itemize}
\item 1687 Isaac Newton stated that the shear stress between layers of a fluid is proportional to the velocity gradient in the direction perpendicular to the layers.
\item 1738 Daniel Bernoulli proved that the gradient of pressure is proportional to the acceleration.
\item 1759 Leonhard Euler  applied Newton's second law of motion to fluid dynamics (no viscosity) and published the {\em Euler equations} of fluid motion.
\item 1807 Jean Baptiste Joseph Fourier presented his  heat conduction equation to the Institut de France.
\item 1822 Claude Navier introduced viscosity in the Euler equation.
\item 1823 Augustin-Louis  Cauchy published his general theory of stress.
\item 1845 George Stokes published the {\em Navier--Stokes equations}.
\item 1872  Ludwig Boltzmann published `Kinetic Theory of Gases' and introduced the {\em Boltzmann equation} for evolution of the particle distribution in the space of  possible positions and momenta.
\item 1879 James Clerk Maxwell used Boltzmann's kinetics to come up with additional terms in the Navier--Stokes equations (the `final equations of motion'), in order to explain the {\em thermal creep} in rarefied gases which cannot be captured by the Navier--Stokes--Fourier equations.
\item 1893 Johannes Diderik van der Waals introduced the thermodynamic theory of capillarity under the hypothesis of a continuous variation of density.
\item 1900 David Hilbert presented his prominent  list of problems. The sixth Hilbert problem includes `the problem of developing mathematically the limiting processes,..., which lead from the atomistic view to the laws of motion of continua'.
\item 1901 Diederik Johannes Korteweg proposed the dynamics equations for fluids with capillarity, based on  van der Waals' approach.
\item 1917 David Enskog defended his thesis in which he gave a systematic derivation of the Navier-Stokes-Fourier equations from the Boltzmann equation and predicted the thermodiffusion effect in the mixtures of gases. The latter was used in early uranium separation technologies.
\item 1932 Steven Chapman and Steven Cowling refined the method of Enskog and wrote the classical treatise `Mathematical theory of non-uniform gases'.
\end{itemize}

Thus, at the beginning of the 20th  century, the main equations we intend to discuss were invented. The Navier--Stokes--Fourier equations and the Boltzmann equation were known and Hilbert considered  reduction from kinetics (`the atomistic view') to fluid mechanics as a crucially important problem.  Van der Waals proposed to include in the energy density a new term, $ \sim c(\nabla \rho)^2$, where $\rho$ is the density and $c$ is the capillarity coefficient \cite{van der Waals1893}. His doctorate student, Korteweg, found how this term affects the motion of fluids, and created the dynamics of fluids with capillarity \cite{Korteweg1901}.

Since then the research tree has branched enormously. The van der Waals capillarity term played an essential role in a number of important achievements: in the Landau theory of phase transitions \cite{Landau1937} and the  Ginzburg--Landau theory of superfluids and superconductors \cite{Ginzburg1950}, in Cahn--Hilliard models of phase separation \cite{Cahn1958} and in the Langer Bar-On and Miller theory of spinodal decomposition \cite{Langer1975}, to name a few. Many new  equations of motion were invented and studied for  various continuous media, and a special discipline, {\em rational thermodynamics}, was developed to manage this world of continuum mechanics models \cite{TruesdellNoll2004,Dmitrienko2011}. In particular, the Korteweg equations were slightly revised by means of rational thermodynamics \cite{Dunn1985}.

The fluid dynamics equations were rigorously derived from Boltzmann's kinetics in some {\em rescaling limits}: if we consider a flow of an almost equilibrium gas with very small space derivatives of velocities and density and then change the time scale (go to `slow time') then the Euler equations hold for this flow with high accuracy. For some scaling procedures, the Navier--Stokes--Fourier equations can be achieved as well. (Exact statements, more details, more results and further references can be found in the book \cite{Saint-RaymondBook}.)

Several families of methods were developed in theoretical physics for deduction of fluid dynamic equations from kinetic equations. We introduce some of them later. They usually work well for the derivation of the `expected' Euler and Navier--Stokes--Fourier equations but the next step, a correction to the Navier--Stokes equations  for more non-equilibrium regimes, was far less successful. Therefore, the most challenging goal of these methods, the correct new post--Navier--Stokes--Fourier equations, remained unachieved.

\subsection{The problem}

In 1946 John von Neumann stated that  computational fluid dynamics would make experimental fluid dynamics obsolete \cite {VonNeumann1946} (see also \cite{Harper2012}).

The computational approach to the problem as discussed by von Neumann has been
developing in many relevant works published meanwhile including e.g. very recent publications \cite{Pikovsly2016} or \cite{Brunton2016}.

\begin{figure}[t]
\begin{centering}
\boxed{\includegraphics[width=0.32 \textwidth]{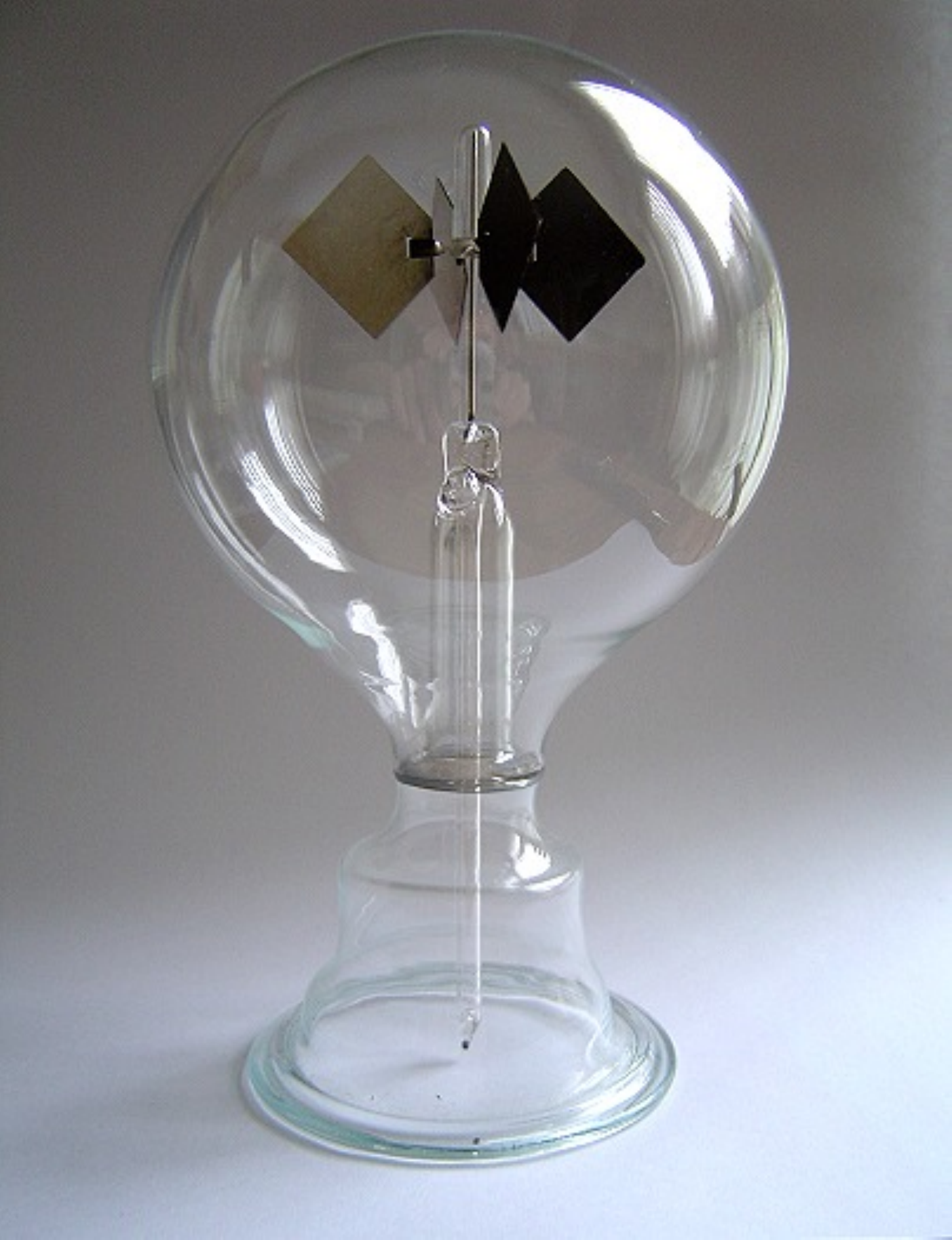}}
\caption {\label{LightMill}{\em Crookes radiometer}. It is made from a glass bulb with low air pressure inside. Within the bulb there is a rotor with several vertical  metal vanes, which are polished on one side and black on the other. When exposed to light the vanes turn, the dark sides retreating from the radiation source and the light sides advancing.  (Public domain picture.)}
\end{centering}
\end{figure}

 But rephrasing the famous quotation of Lord Kelvin \cite{Kelvin1901} we can say that the beauty and clearness of the theory, which explains fluid dynamics  equations as special limits of kinetic equations, is at present obscured by a cloud. Higher-order corrections to the Navier--Stokes--Fourier equations produced from kinetics beyond the degenerated scaled flows demonstrate some non-physical properties and cannot be used without regularization. Moreover, there are relatively simple experiments with rarefied gases (known to Reynolds and Maxwell in 1879 \cite{Reynolds1879,Maxwell1879})  which can  be captured by neither the Navier--Stokes--Fourier equations nor by the higher-order corrections obtained from the Boltzmann equations by regular procedures. These are, for example, effects of thermal creep (or transpiration), i.e. the steady motion of a rarefied gas  induced by a temperature gradient on the boundary of the flow domain, which are demonstrated by the Crookes radiometer (the light mill, Fig.~\ref{LightMill}).

\begin{figure}[t]
\begin{centering}
\boxed{\includegraphics[width= 0.9\textwidth]{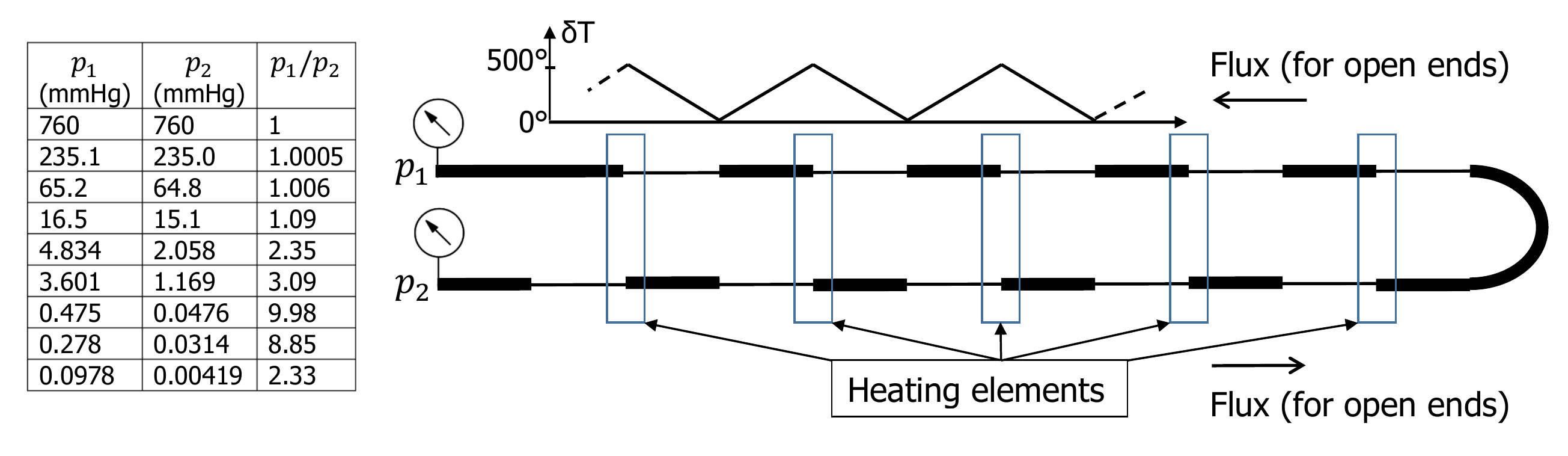}}
\caption{\label{KnudsenPump}{\em Knudsen pump}. Bold lines correspond to thick pipes.  The direction of flux (for open pipe ends) is indicated. For closed pipe ends, $p_1>p_2$. Original Knudsen's experimental data for stationary pressure are in the table. The plot of stationary temperature difference  $\delta T$ from non-heated pipe joints is presented schematically. }
\end{centering}
\end{figure}

Another example of such a system is the Knudsen pump \cite{Knudsen1910}. In Knudsen's experiments,  it was a pipe with alternating narrow (diameter 1/3 mm) and wide (diameter 10 mm) segments  of 5 cm length (Fig.~\ref{KnudsenPump}). Every second pipe joint was heated by a special heating element. Metallic wire was used for heat removal from other joints. The temperature difference between the heated and unheated pipe joints was \SI{500}{\degree}C.  For normal pressure, there was no difference in pressure at the opposite ends of the pipe, whereas for the pressure at one end, $p_1\sim 0.5$ mmHg (65 Pa) the ratio between these pressures is $\sim 10$. Recently, this system was revisited, both by the reproduction of experiments and by theoretical analysis \cite{Sone2000}. The effect was also demonstrated on two joined pipes of different inner diameters, 12 and 24 mm, each 60 mm in length. The joint was heated to about \SI{500}{\degree}C, the ends were cooled by  thick copper plates. The flux from the thinner to the thicker pipe was indicated for the pressure 5-50 Pa. At pressure 100 Pa (and above) there was no motion observed in this system \cite{Sone2000}.

Of course, these effects do not contradict the kinetic theory and can be explained in the framework of Boltzmann's kinetics and some of its simplifications \cite{Ohwada1989,Loyalka}. The problem is in the appropriate continuum mechanics model, which is produced systematically from kinetics and explains  thermal creeps. The prophecy of von Neumann is unfulfilled  not only because the computational tools are insufficient but also because the proper equations for some situations remain unknown.

\subsection{Solutions proposed: Capillarity of ideal gas}

Van der Waals introduced  capillarity energy for non-ideal multiphase fluids \cite{van der Waals1893,Rowlinson2013}. However, the terms which look similar to Korteweg's stress tensor have been recognized in the Chapman--Enskog expansion for non-equilibrium ideal gas many times. Everyone can compare the Korteweg stress (equation (\ref{Kstress}) below) with  Burnett's equations, the first post-Navier-Stokes correction derived from the Boltzmann equation (see, for example, the Chapman and Cowling book \cite{Chapman}).

Nevertheless, it was not before the exact summation of the entire Chapman--Enskog series was found \cite{GKJETP91,GKPRL96,JPA00,KGAnPh2002} that the following rule was discovered  \cite{SlemQuaterly2012,SlemCAMWA2013}:
{\em Chapman--Enskog $\Rightarrow$  Viscosity + Capillarity.}

The problem with Burnett's equations is, among others,  in the wrong {\it sign} of the higher-order correction terms leading to a parasite instability of short waves (Bobylev's instability \cite{Bobylev1982}). Maxwell, in his attempt to describe thermal creep, encountered the wrong sign of the effect (according to Burnett's equations the light mill rotates in the direction opposite to the experimentally observed rotation).

The exact solutions demonstrate that the van der Waals capillarity emerges in the non-equilibrium {\it ideal} gas. Therefore, Korteweg's equations can be a candidate for the  post--Navier--Stokes--Fourier equations. The fact that the van der Waals capillarity energy emerges from a rigorous analytic solution of the reduction problem for kinetic models of ideal gas was unexpected.

On the other hand, the exact solutions demonstrate saturation of dissipation: the attenuation rate of the acoustic waves has a finite limit when the wave length tends to zero \cite{GKJETP91,KGAnPh2002}. In that respect, the original Korteweg's equation may be not sufficient: for them, the attenuation rate grows as (frequency)$^{2/3}$ (for high frequency, see (\ref{EhrenRoots}) and Fig.~\ref{SpectraKortewEhren}b below). Two different hypotheses seem to be possible at present:
\begin{itemize}
\item For the genuine Boltzmann equation with realistic interaction between the particles, the asymptotic of the attenuation rate of acoustic waves  is the same as for the Korteweg equations;
\item The proper hydrodynamics for non-equilibrium gas should include the mixed derivatives, which look like the operator $(1-\alpha \Delta) \upartial_t $ in the left hand side of the equations (in the simplest form), where $\Delta$ is the Laplace operator.
\end{itemize}
There are some arguments in favor of the Korteweg asymptotic reported very recently \cite{HuangYang2016}. Nevertheless, the second hypothesis is strongly supported by the exactly solved problem, as we demonstrate below.

\subsection{Capillarity of ideal gas in words and numbers}

In high school textbooks and popular science literature, capillarity is the ability of a liquid to flow in narrow gaps without the assistance of, and in opposition to, external forces such as external pressure or gravity. In scientific literature, the term `capillarity’ is used in wider sense. For example, the pressure difference across the curve interface is termed the `capillary pressure’.
Gibbs theory of capillarity \cite{Gibbs1877}   carries the title “Influence of Surfaces of Discontinuity upon the Equilibrium of Heterogeneous Masses.” This title includes two important ideas: {\em equilibrium} and {\em discontinuity}. On the contrary, van der Waals  \cite{van der Waals1893,Rowlinson2013} proposed the theory of {\em continuous} transition layers which was developed later in the {\em dynamic}  phase field theory and numerics. In the limit of the thin equilibrium film, the van der Waals theory gives results very similar to the Gibbs theory.

Gibbs studied energy and equilibria of interface surfaces. We can call them infinitely `thin films’, films without thickness. Van der Waals and, later, his student Korteweg studied continuous media with energy dependent on density gradients. We can call the van der Waals transition layers `thick films’, they have non-negligible thickness and volume. Now, we aim to demonstrate that even very simplified piecewise continuous approximation of a thick film can give useful insights.

\begin{figure}[t]
\begin{centering}
\boxed{\includegraphics[width= 0.3\textwidth]{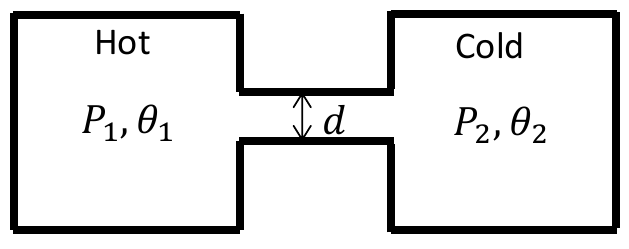}}
\caption{\label{Chambers}Two chambers connected by a capillary tube. }
\end{centering}
\end{figure}

Consider two chambers with fluid connected by a capillary tube (Fig.~\ref{Chambers}). Assume that a  system of heaters and coolers keeps the temperatures, $\theta_1>\theta_2$,  in these chambers  constant in time. In a steady state, there is no fluid flux through the tube but a heat flux exists and so this is not a thermodynamic equilibrium. Two limit cases for gases are very well known:
\begin{itemize}
\item {\em The Navier--Stokes limit}, the mean  free  path $\lambda \ll d$. In this limit zero flux implies zero pressure difference. In the steady state $P_1=P_2$;
\item {\em The Knudsen limit}  $\lambda \gg d$, the flux in the capillary tube is collisionless. The gas  flux from the hot chamber to the cold chamber is $const \times n_1 \bar{v}_1$ and the inverse flux is
$const \times n_2 \bar{v}_2$ with the same $const$ (we presume that the capillary tube is symmetric). Here, $n_i$ is density, the ideal gas equation gives $n_i k\theta_i=P_i$, $\bar{v}_i$ is the thermal velocity, proportional to $\sqrt{k_{\rm B}\theta_i}$, and $k_{\rm B}$ is the Boltzmann constant. The zero flux condition gives
\begin{equation}\label{KnudsenLimit}
\frac{P_1}{\sqrt{\theta_1}}=\frac{P_2}{\sqrt{\theta_2}}\;\;\mbox{or}\;\;  \frac{P_2}{P_1}=\sqrt{\frac{\theta_2}{\theta_1}}.
\end{equation}
\end{itemize}
In the Knudsen limit, we see how  the steady state difference between temperatures produces the difference between pressures. This is called the {\em thermomechanical effect}, and the effect of the capillary flux induced by the temperature gradient is called {\em thermoosmosis}. The ratio $P_2/P_1$ is called the {\em thermal transpiration ratio}.

The thermomechanical effect and thermoosmosis were discovered  in the 20th century for various fluids, from rarefied gases to superfluid  liquids and water \cite{Derjaguin1941,Denbigh1948}. Now, they are considered as a promising driving mechanism for micro- or nanoscale engines \cite{Semenov2015}.
There are many approaches to the theory of these effects based on non-equilibrium thermodynamics \cite{Rastogi2007}. All of them include unknown parameters, which should be found experimentally. Here we will demonstrate two simple models based on capillarity (perhaps, the simplest models). First of all, assume that the thermomechanical effect is produced by a thin layer at the wall of the capillary. The property of this surface depends on the thermodynamic conditions and on the material of the wall and the fluid.

\begin{figure}[t]
\begin{centering}
\boxed{\includegraphics[width= 0.4\textwidth]{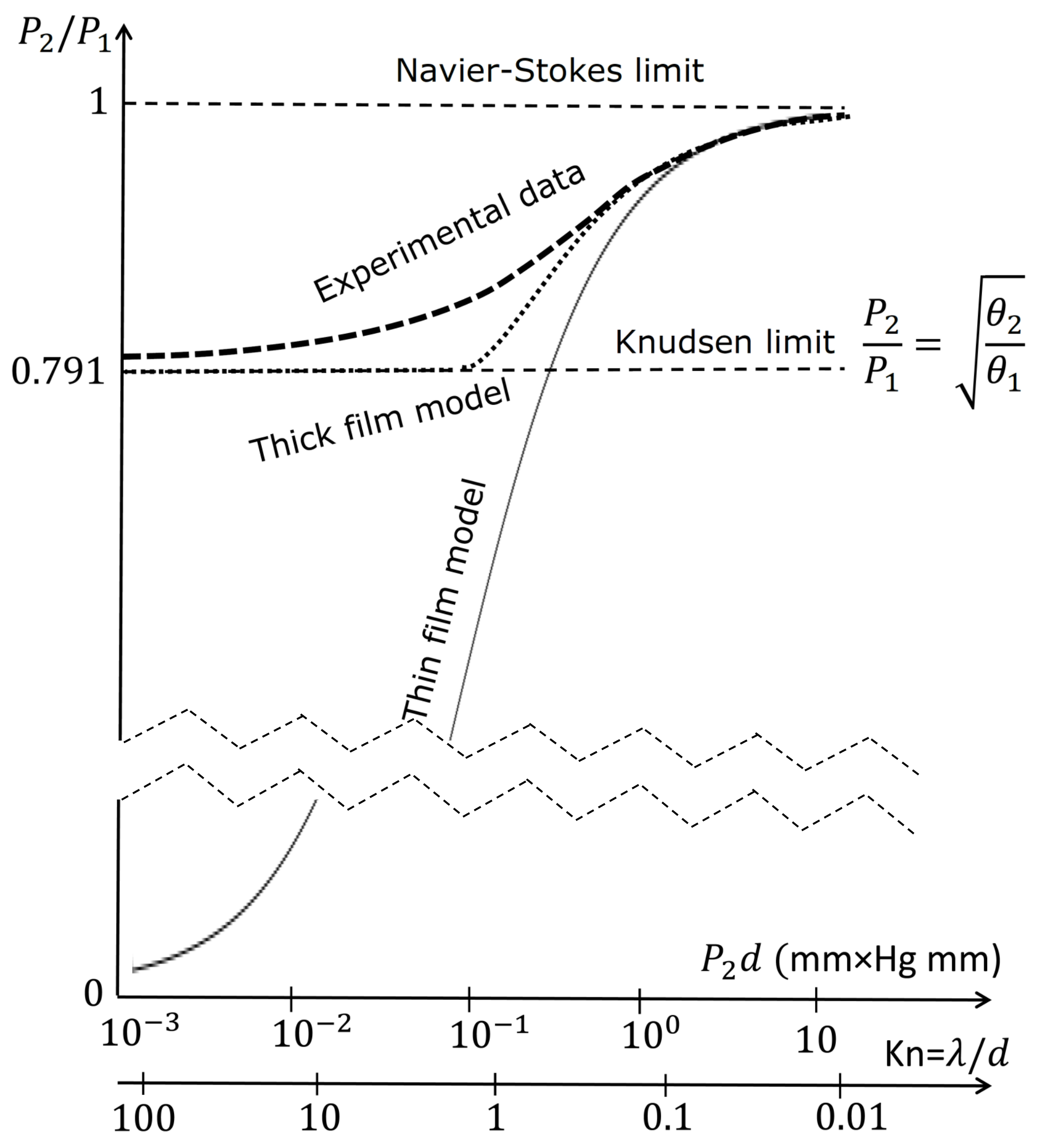}}
\caption{\label{Thermomechanic}Thermal transpiration values $P_2/P_1$ of hydrogen. The simplest thin film  capillarity model (solid line) and the thick film model (bold dotted line) versus experimental data (bold dashed line).  $P_1$  is the pressure at the `hot end', $T_1=473\pm 1 {\rm K}$,  $ P_2$ is the pressure at the `cold end', $T_2=295\pm1 {\rm K}$ (room
temperature). Experimental data are taken from \cite{Takaishi1963}. The simplest capillarity model gives: $P_2/P_1=P_2 d/(P_2d+1/15)$ (fitted to data in the interval $ 0.01<{\rm Kn}<0.1$). Kn (the ratio of the mean free path $\lambda$ to the diameter $d$) is evaluated at the cold end.  We used the collisional diameter of H$_2$, $D_{\rm H2}\approx 2.3 \times 10^{-10} {\rm m}$ estimated  from viscosity data using mean free path theory. In the analysis of molecular seeds a larger diameter is used usually, $D_{\rm H2} \approx 2.89 \times 10^{-10} {\rm m}$ \cite{Breck1974}. In the thick film model $a\approx 0.25$ for the experimental data presented.}
\end{centering}
\end{figure}

In the Gibbs approach \cite{Gibbs1877} we operate with a thin film  (a surface). The force in the thermomechanical effect should be proportional to the length of the section of this surface, that is the circle, $F=\gamma \pi d$, and the pressure difference should be (Fig.~\ref{Thermomechanic})  $$P_1-P_2=\frac{F}{\frac{1}{4}\pi d^2}=\frac{4\gamma}{d}\;\; \mbox{and} \;\; \frac{P_2}{P_1}=\frac{P_2 d}{P_2d+4 \gamma}.$$

This model includes one unknown coefficient $\gamma$, which can be easily evaluated from the empiric curves (Fig.~\ref{Thermomechanic}). Of course, this simplest model belongs to  Type 5 of  Peierls' classification \cite{Peierls1980}:  `Instructive model (No quantitative justifications but gives insight)' or even to Type 6: `Analogy (Only some features in common)'. This model works well in comparison with experiment  (Fig.~\ref{Thermomechanic})  for not very high Knudsen numbers ${\rm Kn }<0.1$, gives the proper Navier--Stokes limit,  but has one fundamental defect: $P_2/P_1 \to 0$ when $P_2d \to 0$ (or, which is the same, Kn$\to \infty$).

To improve the model, let us follow the van der Waals capillarity idea. Consider a `thick film' near the border with thickness $\varsigma$, if $d>2\varsigma$.  If $d<2\varsigma$ then the thick surface fills the whole capillary.  Assume that the thickness is proportional to the mean free path, $\varsigma=a\lambda$. The force of the thermomechanical effect in this model is  proportional to the section of the surface orthogonal to the axis of the capillary (to the surface area), and to the particle density (empty space does not produce force): $F=\kappa A n$, where $A$ is the area of the section. The coefficient $\kappa$ has the physical sense of `energy per particle' and describes the energetic difference between the boundary layer of thickness $\varsigma$ and the rest of the volume of the capillary tube.  Recall that $n\lambda=(\sqrt{2} \pi D^2)^{-1}$, where $D$ is the collision diameter of molecules, and $P=nk_{\rm B}\theta$. Evaluate $\lambda$ at the cold end. After simple algebra we get  the  formula for the transpiration ratio in the thick film approximation:
\begin{equation*}
\frac{P_2}{P_1}=\left\{\begin{array}{cl}
 \frac{P_2 d}{P_2d+4\kappa a (\sqrt{2} \pi D^2)^{-1} (1-a {\rm Kn})} & \mbox{if  }a {\rm Kn}<\frac{1}{2}, \\
 \frac{1}{1+\kappa (k_{\rm B}\theta_2)^{-1}}& \mbox{if  } a{\rm Kn}\geq \frac{1}{2}. \end{array} \right.
\end{equation*}
Here, $\kappa$ is still an unknown coefficient. The thick film fills the whole capillary at Kn=1/(2a).
It is convenient to gather all the coefficients into two unknown numbers: coefficient $\alpha$ (a function of temperature, which can be easily extracted from the Knudsen asymptotic) and the ratio of the film thickness to the mean free path, $a=\varsigma/\lambda$.  We express the thermal transpiration ratio through the Knudsen number:
\begin{equation}
\frac{P_2}{P_1}=\left\{\begin{array}{cl}
 \frac{1}{1+\alpha {\rm Kn} (1-a {\rm Kn})} & \mbox{if  } a{\rm Kn}<\frac{1}{2}, \\
 \frac{1}{1+{\alpha}/(4a)}& \mbox{if  } a{\rm Kn}\geq \frac{1}{2}. \end{array} \right.
\end{equation}

The thick surface model has the proper high Kn and low Kn asymptotics and can be used for the understanding the data. Of course, it is not far from the simple dimensional analysis and  belongs to Type 5 or 6 of Peierls' classification. It does not reveal the intrinsic physical mechanism of the thermomechanic effect and the thermoosmosis. We just hypothesized that these effects are connected to the energy (enthalpy) density in the layers near the interface between gas and capillary and to the transport in these layers (thermoosmosis slip). The models can be improved if we account carefully for variations of the films along the tube.  Solving the Korteweg equations instead of postulating of the thick surface can provide the further improvements, but this next step requires more technical effort. Some calculations of boundary layers and transpiration were performed recently on the  basis of Korteweg's equation \cite{KimLeeSlemrod}. The problem is not in an accurate calculation of these effects. The kinetic equations or even direct molecular simulation do this job quite well. The question is about an appropriate continuum model that describes the bulk motion, the thermal transpiration and the other similar effects in a unified  way.

In any case, already very simple models demonstrate that the transport phenomena of the rarefied gas near the interfaces may be described using the idea of capillarity. We might imagine that the light mill is a close relative of the soap boat \cite{SoapBoat2013} because both are moving by capillarity effects.

\subsection{The structure of the paper}

The goal of this paper is to highlight the origin of the capillarity of ideal gas.
For this purpose, we give a comprehensive introduction into modern theory of dynamic model reduction from kinetics to fluid dynamics and analyze approximate and recently found exact solutions of this problem.

In the next Section, we start the unhurried introduction of the main equations: from the conservation of mass and Cauchy equations to the Navier--Stokes--Fourier equations. In Section~\ref{Sec:KortewegStress}, Korteweg's stress tensor is described. After that, we discuss a simple `mean free path' theory for Korteweg's stress. To that end, the idea of Ehrenfests'
coarse-graining is utilized: we take the collisionless Knudsen gas with periodic equilibration to the local equilibrium and analyze the post-Navier--Stokes equation derived by this approach.

After that, a short but comprehensive explanation of the Chapman--Enskog procedure is presented. We introduce the invariance equation in Section~\ref{Sec:invariance} and construct the Chapman--Enskog as a formal solution to this equation (Section~\ref{Sec:Ch-E}). The geometric language allows us to avoid the usual bulky calculations while describing the main constructions.

Analysis of exactly solvable problem of reduction allows us to prove that the capillarity effects in ideal gas is not a by-product of  special  approximations and series expansions but appears as a direct consequence of kinetic equations. In Section~\ref{Sec:CapIdGas} the energy formula for
hydrodynamics of non-equilibrium ideal gas is proved. It includes the pseudodifferential capillarity
term, which coincides with the van der Waals energy of capillarity in first approximation. Finally,  hypotheses are discussed about the plausible forms of non-equilibrium fluid mechanics.

\section{Fluid dynamics equations}
\subsection{Hydrodynamic fields}
Consider the classical fluid, which is defined by the {\em hydrodynamic fields}: density $\rho$ (scalar), velocity $\boldsymbol{u}$ (vector), and  specific internal energy $e$, that is internal energy per unit of mass (scalar).

\subsection{Conservation of mass}
\begin{equation}\label{ConservationMass}\boxed{
\frac{\upartial \rho}{\upartial t}+\sum_j \frac{\upartial (\rho u_j)}{\upartial x_j}=0.}\footnote{The most important equations are boxed.}
\end{equation}

\subsection{Cauchy momentum equations}
\begin{equation}\label{ConservationMomentum}\boxed{
\frac{\upartial (\rho u_i)}{\upartial t}+ \sum_j\frac{\upartial (\rho u_i u_j)}{\upartial x_j}=\sum_j \frac{\upartial \sigma_{ij}}{\upartial x_j} +f_i.}
\end{equation}
where $\boldsymbol{f}=(f_i)$ is the body force density, $\sigma_{ii}$ ($i=1,2,3$) are normal stresses, and  $\sigma_{ij}$ ($i\neq j$) are shear stresses. The pressure $p$ is $$p=-\frac{1}{3}{\rm tr }\boldsymbol{\sigma} =-\frac{1}{3}\sum_i \sigma_{ii}.$$

The Cauchy stress tensor $\boldsymbol{\sigma}=(\sigma_{ij})$ ($i,j=1,2,3$)  describes the so-called {\em contact force}: Let for a small element of a body surface with area $\Delta S$ and outer normal vector $\boldsymbol{n}=(n_i)$ the stress tensor $\boldsymbol{\sigma}$ be given. Then the force applied  {\em to the body} through this surface fragment is $$\Delta \boldsymbol{F}=\boldsymbol{\sigma} \cdot \boldsymbol{n}\Delta S, \mbox{  that is,  }F_i=\Delta S \sum_j \sigma_{ij} n_j.$$
The stress tensor is symmetric (to provide conservation of angular momentum), $\sigma_{ij}=\sigma_{ji}$.

The momentum equation (\ref{ConservationMomentum}) is the second Newton law written for a small brick of the material: Imagine a small cubic brick (a `parcel') with soft but impenetrable boundaries and edges parallel to the coordinate axes, which moves with the material. The outer normals of the opposite faces differ by sign only, therefore the sum of the contact forces from all faces is   $\sum_j \frac{\upartial \sigma_{ij}}{\upartial x_j} \Delta L^3 + o(\Delta L^3)$, where $\Delta L$ is the edge of the brick  (use the Taylor formula for $\sigma_{ij}$). The possibility to represent the motion of continuum as the flight of many infinitesimal parcels with soft deformable but impenetrable boundaries is in  the essence of the mechanics of materials.

\subsection{Energy equation}
If the material body $\Omega$ experiences infinitisemal {\em displacement} $\boldsymbol{x} \mapsto \boldsymbol{x}+\delta \boldsymbol{r(x)}$, which vanishes with derivatives on the boundary,  then the work of the contact forces is (use the Stokes formula for integration by parts):
\begin{equation}\label{work}
\delta W=\int_{\Omega} \sum_{i,j} \frac{\upartial \sigma_{ij}}{\upartial x_j} \delta r_i  \, \D ^3x=
-\int_{\Omega} \sum_{i,j} \sigma_{ij}\frac{\upartial \delta r_i}{ \upartial x_j} \, \D ^3x .
\end{equation}
Take into account symmetry of $\sigma_{ij}$ and find that $\delta W$ depends on the symmetric part of  ${\upartial \delta r_i}/{ \upartial x_j}$:
$$\delta W=-\frac{1}{2}\int_{\Omega} \sum_{i,j} \sigma_{ij}\left(\frac{\upartial \delta r_i}{ \upartial x_j} +\frac{\upartial \delta r_j}{ \upartial x_i}\right)  \, \D ^3x .$$
The power is the work per time. For a moving of continuum media, the power produced by the contact forces is
\begin{equation}\label{power}
P=-\int_{\Omega} \sum_{i,j} \sigma_{ij}\frac{\upartial u_i}{ \upartial x_j} d^3x=-\frac{1}{2}\int_{\Omega} \sum_{i,j} \sigma_{ij}\left(\frac{\upartial u_i}{ \upartial x_j} +\frac{\upartial u_j}{ \upartial x_i}\right)  \, \D ^3x .
\end{equation}

According to the Cauchy equation and the formula for power we can write the conservation of energy
\begin{equation}\label{energy}\boxed{
\frac{\upartial  (\rho \epsilon)}{\upartial t}+\sum_j \frac{\upartial (\rho \epsilon u_j)}{\upartial x_j}= \sum_{ij}\frac{\upartial (\sigma_{ij} u_i)}{ \upartial x_j} + \sum_i f_i u_i-\sum_i \frac{\upartial q_i}{\upartial x_i},}
\end{equation}
where $\epsilon$ is the energy density per unit mass: $\epsilon=e+\frac{1}{2}u^2$, $e$ is the {\em specific internal energy}, $u^2=\sum_i u_i^2$, and $\boldsymbol{q}=(q_i)$ is the vector of {\em heat flux}, which appears because of thermal conductivity. For the perfect (monoatomic) gas, $e=\frac{3R}{2M} \theta$, where $R$ is the ideal gas constant, $M$ is the molar mass and $\theta$ is the absolute temperature.

\subsection{Stress tensor and heat flux for ideal fluid: the Navier--Stokes--Fourier equations}

Cauchy equations are simple and nice, and can be applied to many materials (and even beyond the mechanics of materials). The question is: how can we express the stress tensor and the heat flux through the hydrodynamic fields:  density $\rho$, velocity $\boldsymbol{u}$, and  specific internal energy $e$?

The simplest answer is given by the Navier--Stokes stress tensor and Fourier heat flux:
\begin{equation}\label{NSFstress}\boxed{
\begin{split}
\sigma_{ij}&=\sigma_{ij}^E+\sigma_{ij}^V; \\
\sigma_{ij}^E&=-p\delta_{ij};\\
\sigma_{ij}^V&=\lambda \delta_{ij} \sum_k \frac{\upartial u_k}{\upartial x_k} + \mu \left(\frac{\upartial u_i}{\upartial x_j}+\frac{\upartial u_j}{\upartial x_i}\right) ;\\
q_i&=-\kappa \frac{\upartial \theta}{\upartial x_i}.
\end{split}}
\end{equation}
Here (and in what follows), $\delta_{ij}$ is Kronecker's delta, $\sigma_{ij}^E$, $\sigma_{ij}^V$ are elastic and viscous contributions to the Cauchy stress tensor, the coefficients $\lambda$, $ \mu$ and $\kappa$ depend on density and temperature. The thermodynamic pressure $p$ in the elastic stress is defined through the equation of state. For example, for ideal gas $p=nR\theta=\rho \theta R/M$, where $n$ is the molar density  ($n=\rho/M$). The standard assumption is that the viscous tensor has zero trace and, therefore, $\lambda=-\frac{2}{3}\mu$. Strictly speaking, this is just an assumption and the mechanical pressure should not necessarily coincide with the thermodynamic pressure and may have a viscous component.

With the {\em constitutive equations} (\ref{NSFstress}) we can immediately write the Navier--Stokes--Fourier equations for the density, velocity and temperature of ideal gas ($p=\frac{R}{M}\theta $, $e=\frac{3R}{2M}\theta $):
\begin{equation}\label{NSFeq}
\begin{split}
&\frac{\upartial \rho}{\upartial t}+\sum_j \frac{\upartial (\rho u_j)}{\upartial x_j}=0;\\
&\frac{\upartial (\rho u_i)}{\upartial t}+ \sum_j\frac{\upartial (\rho u_i u_j)}{\upartial x_j}+\frac{R}{M}\frac{\upartial (\rho \theta)}{\upartial x_i}\\&=\frac{\upartial}{\upartial x_i}\left(\lambda  \sum_k \frac{\upartial u_k}{\upartial x_k}\right)+
 \sum_j \frac{\upartial}{\upartial x_j} \left[\mu \left(\frac{\upartial u_i}{\upartial x_j}+\frac{\upartial u_j}{\upartial x_i}\right)\right]+f_i;\\
&\frac{\upartial  (\rho [\frac{3R}{2M}\theta+\frac{1}{2}u^2])}{\upartial t}+\sum_j \frac{\upartial (\rho [\frac{3R}{2M}\theta+\frac{1}{2}u^2] u_j)}{\upartial x_j}\\&+\rho \theta \frac{R}{M}\sum_i \frac{\upartial u_i}{\upartial x_i} =\lambda \left(\sum_i \frac{\upartial u_i}{\upartial x_i}\right)^2\\&+\frac{\mu}{2} \sum_{ij}\left(\frac{\upartial  u_i}{ \upartial x_j}+\frac{\upartial  u_j}{ \upartial x_i}\right)^2 + \sum_i f_i u_i + \sum_i \frac{\upartial}{\upartial x_i}\left(\kappa \frac{\upartial \theta}{\upartial x_i}\right).
\end{split}
\end{equation}
The power produced by the viscous stress is
$$P=-\lambda \left(\sum_i \frac{\upartial u_i}{\upartial x_i}\right)^2-\frac{\mu}{2} \sum_{ij}\left(\frac{\upartial  u_i}{ \upartial x_j}+\frac{\upartial  u_j}{ \upartial x_i}\right)^2. $$
If $\mu>0$ and $3\lambda+2\mu\geq 0$ then $P\leq 0$ (i.e. viscosity is friction, indeed).

According to the model classification proposed by R. Peierls \cite{Peierls1980}, the Navier--Stokes viscous stress tensor is a typical example of the {\em linear response models} together with  Ohm's and Hooke's laws, Fourier's law and many others. `This refers to a situation in which one is
interested, by definition, in the response of a system to some parameter in the limit in
which this parameter may be treated as infinitesimal.'

\subsection{Fluids with capillarity: van der Waals  energy and Korteweg stress tensor \label{Sec:KortewegStress}}

In 1877, Gibbs published the theory of capillarity based on the idea of surfaces of discontinuity \cite{Gibbs1877}. He introduced and studied thermodynamics of two-dimensional objects -- surfaces. On the contrary, van der Waals proposed the theory of capillarity using the hypothesis that density of the body varies continuously at and near the transition layer  and the energy depends on the gradient of density \cite{van der Waals1893} (just square of gradient is added). This is an energetic `penalty' for high gradients, which does not allow discontinuities to appear and defines the thickness and the energy of the layer between phases. Modern  `phase field' approaches are also based on this idea \cite{Steinbach1996,Antanovskii1995}.  After van der Waals, the standard approach to the continuous theory of capillarity is: represent the Gibbs free energy of a body $\Omega$ as a sum,  $G=G_0+G_K$, where
\begin{equation}\label{Wenergy}
G_0=\int_{\Omega} g_0(\rho,\theta) \, \D ^3x;\;\;G_K=\int_{\Omega} K(\rho,\theta)\sum_i \left(\frac{\upartial \rho}{\upartial x_i}\right)^2 \, \D ^3 x.
\end{equation}
 Here, $K(\rho,\theta)$ is the capillarity coefficient.
Korteweg found the corresponding $G_K$ addition to the stress tensor:
\begin{equation}\label{Kstress}\boxed{
\sigma^K_{ij}=\rho \left[\sum_k \frac{\upartial}{\upartial x_k}\left( K(\rho,\theta)\frac{\upartial \rho}{\upartial x_k}\right)\right]\delta_{ij}-  K(\rho,\theta)\frac{\upartial \rho}{\upartial x_i}\frac{\upartial \rho}{\upartial x_j}.}
\end{equation}
Detailed analysis of this stress tensor and its various modifications from the rational thermodynamics point of view is performed by Dunn and Serrin \cite{Dunn1985}.
Thermodynamics requires to complement the Korteweg stress tensor by the additional contributions to the internal energy and the heat flux:
\begin{equation}\label{Kheat}\boxed{
e^K= \frac{1}{2\rho}\left(K-\theta\frac{\upartial K}{\upartial \theta}\right) \sum_i \left(\frac{\upartial \rho}{\upartial x_i}\right)^2; \;\; q^K_i=K(\rho,\theta)\rho \frac{\upartial \rho}{\upartial x_i}\sum_j\frac{\upartial u_j}{\upartial x_j}.}
\end{equation}
To modify the Navier--Stokes equation we should just add the Korteweg stress to the elastic and viscous stresses, and also add the capillarity contributions to the energy and to the heat flux
\begin{equation}\label{NSFKstress}
\sigma_{ij}=\sigma_{ij}^E+\sigma_{ij}^V+\sigma^K_{ij}; \;\; \epsilon=e+\frac{1}{2}u^2+e^K; \;\; q_i=-\kappa \frac{\upartial \theta}{\upartial x_i}+q^K_i.
\end{equation}
 With these additions, the Navier--Stokes--Fourier equations for ideal gas (\ref{NSFeq}) turn into {\em Korteweg equations} for gas with capillarity. Now we have to understand how it may happen that the ideal gas gains capillarity.

\section{Kinetics}

\subsection{Stress tensor for collisionless gas}

Collisionless gas consists of particles which are moving without collisions between them. It is described by the one-particle distribution function $f(\boldsymbol{x},\boldsymbol{v},t)$, where $\boldsymbol{x}$ is  particle's position, $ \boldsymbol{v}$ is particle's velocity, and $t$ is time. The integral of the density over a  domain $\Omega$ in the 6-dimensional space of positions and velocities is the number of particles in  $\Omega$: $\int_{\Omega} f(\boldsymbol{x},\boldsymbol{v},t) d^3x \, \D ^3v=N(t)$.

Time evolution of $f(\boldsymbol{x},\boldsymbol{v},t)$ is given by a simple explicit expression:
\begin{equation}\label{freeflight}\boxed{
f(\boldsymbol{x},\boldsymbol{v},t)=f(\boldsymbol{x}-\boldsymbol{v}t,\boldsymbol{v},0). }
\end{equation}
It satisfies the {\em advection equation}
\begin{equation}\label{advection}
\frac{\upartial f(\boldsymbol{x},\boldsymbol{v},t)}{\upartial t}  + \sum_i v_i \frac{\upartial f(\boldsymbol{x},\boldsymbol{v},t)}{\upartial x_i}=0.
\end{equation}
The hydrodynamic  variables are (for particles of unit mass):
\begin{equation}\label{moments}
\rho=\int f(\boldsymbol{x},\boldsymbol{v},t) \, \D ^3 v; \;
\rho \boldsymbol{u}=\int \boldsymbol{v} f(\boldsymbol{x},\boldsymbol{v},t)  \, \D ^3 v; \;
\rho \epsilon=\frac{1}{2}\int v^2 f(\boldsymbol{x},\boldsymbol{v},t)  \, \D ^3 v.
\end{equation}
Integration of the advection equation (\ref{advection}) in $v$ gives the conservation of mass equation (\ref{ConservationMass}). If we multiply the advection equation by $v_i$ then the integration  gives the momentum equation and after multiplication with $\frac{1}{2}v^2$ the integration produces the energy equation:
\begin{equation}
\frac{\upartial \rho}{\upartial t}+\sum_j \frac{\upartial (\rho u_j)}{\upartial x_j}=0; \;
\frac{\upartial \rho u_i}{\upartial t}+\sum_j \frac{\upartial \Pi_{ij}}{\upartial x_j}=0;\;
\frac{\upartial \rho \epsilon }{\upartial t}+\sum_j \frac{\upartial Q_j}{\upartial x_j}=0.
\end{equation}
Here, tensor $\Pi_{ij}$ of {\em momentum flux} and vector $Q_i$ of {\em energy flux} are defined as moments of the distribution $f$ :
\begin{equation}
\Pi_{ij}=\int v_iv_j f(\boldsymbol{x},\boldsymbol{v},t)  \, \D ^3 v; \;\;Q_i=\frac{1}{2}\int v_i v^2 f(\boldsymbol{x},\boldsymbol{v},t)  \, \D ^3 v.
\end{equation}
It is convenient to separate the flux with the mean flow velocity $\boldsymbol{u}$ from other forms of transport. For this purpose, the central moments of $f$ are convenient. Simple algebra gives:
\begin{eqnarray}
&&\rho e=\frac{1}{2}\int (\boldsymbol{v-u})^2 f(\boldsymbol{x},{v},t)  \, \D ^3 v, \;\;  \rho \epsilon=\rho e + \frac{1}{2}\rho u^2;\; \\
&&\sigma_{ij}=-\int (v_i-u_i)(v_j-u_j) f(\boldsymbol{x},\boldsymbol{v},t)  \, \D ^3 v, \;\; \Pi_{ij}=u_iu_j-\sigma_{ij};\\
&&q_i=\frac{1}{2}\int (v_i-u_i) (\boldsymbol{v-u})^2 f(\boldsymbol{x},{v},t)  \, \D ^3 v,\;\; Q_i=q_i-\sum_j \sigma_{ij}u_j+u_i \rho \epsilon
\end{eqnarray}
(the minus sign in the definition of $\sigma$ appears just to provide similarity to Cauchy stress).
Recall that here for simplicity, we take unit mass of the particles and measure $\rho$ in number of particles per unit volume. To return to the standard absolute temperature $\theta$ we have to notice that $e=\frac{3}{2}k_{\rm B}\theta$, where $k_{\rm B}$ is Boltzmann's constant.

With usage of central moments, the transport equations look very similar to the 	qCauchy and energy equations:
\begin{equation}\label{FreeFlightCauchi}
\begin{split}
&\frac{\upartial \rho}{\upartial t}+\sum_j \frac{\upartial (\rho u_i)}{\upartial x_i}=0; \\
&\frac{\upartial  (\rho u_i)}{\upartial t}+\sum_j \frac{\upartial( \rho u_i u_j)}{\upartial x_j}=\sum_j \frac{\upartial \sigma_{ij}}{\upartial x_j}; \\
&\frac{\upartial (\rho \epsilon)}{\upartial t}+\sum_j \frac{\upartial(\rho \epsilon u_j)}{\upartial x_j}+\sum_j \frac{\upartial q_j}{\upartial x_j}=\sum_{ij}\frac{\upartial (\sigma_{ij} u_i)}{ \upartial x_j}.
\end{split}
\end{equation}
The standard definition of pressure gives:
\begin{equation}\label{PressColless}
p=-\frac{1}{3}\sum_i \sigma_{ii}=\frac{1}{3}\int (\boldsymbol{v}-\boldsymbol{u})^2 f(\boldsymbol{x},\boldsymbol{v},t)  \, \D ^3 v=\frac{2}{3}\rho e.
\end{equation}

Particles of collisionless gas do not interact and there are no forces. Motion of such a gas cannot be represented as motion of continuum. Nevertheless, the Cauchy stress tensor can be defined through the momentum flux and the Cauchy transport equation holds (\ref{FreeFlightCauchi}).

The elastic part (pressure) of the stress tensor for collisionless gas has the very common form (\ref{PressColless}). It is not surprising that the rest of the stress tensor does not satisfy the Navier--Stokes constitutive relations (\ref{NSFstress}). Nevertheless,  Ehrenfests' idea of coarse-graining  allows us to deduce, step by step, the Euler equations, the Navier--Stokes--Fourier equations and  the Korteweg equations for fluid motion from the simple model of collisionless gas.

\subsection{Ehrenfests' coarse-graining: from collisionless gas to Korteweg fluid dynamics through coarse-graining}

In 1911, Paul and Tanya Ehrenfest in their paper for the scientific Encyclopedia \cite{Ehrenfests1911} discussed emergence of irreversibility and introduced a special operation, that of {\em  coarse-graining}. This operation transforms a probability density in phase space into a `coarse-grained' density. That is a piece-wise constant function, a result of density averaging in cells. The size of cells is assumed to be small, but finite, and does not tend to zero. The coarse-graining (`shaking') models thermalisation of the system by uncontrollable and weak interaction with  surroundings.

\begin{figure}
\begin{centering}
\boxed{\includegraphics[width=0.5 \textwidth]{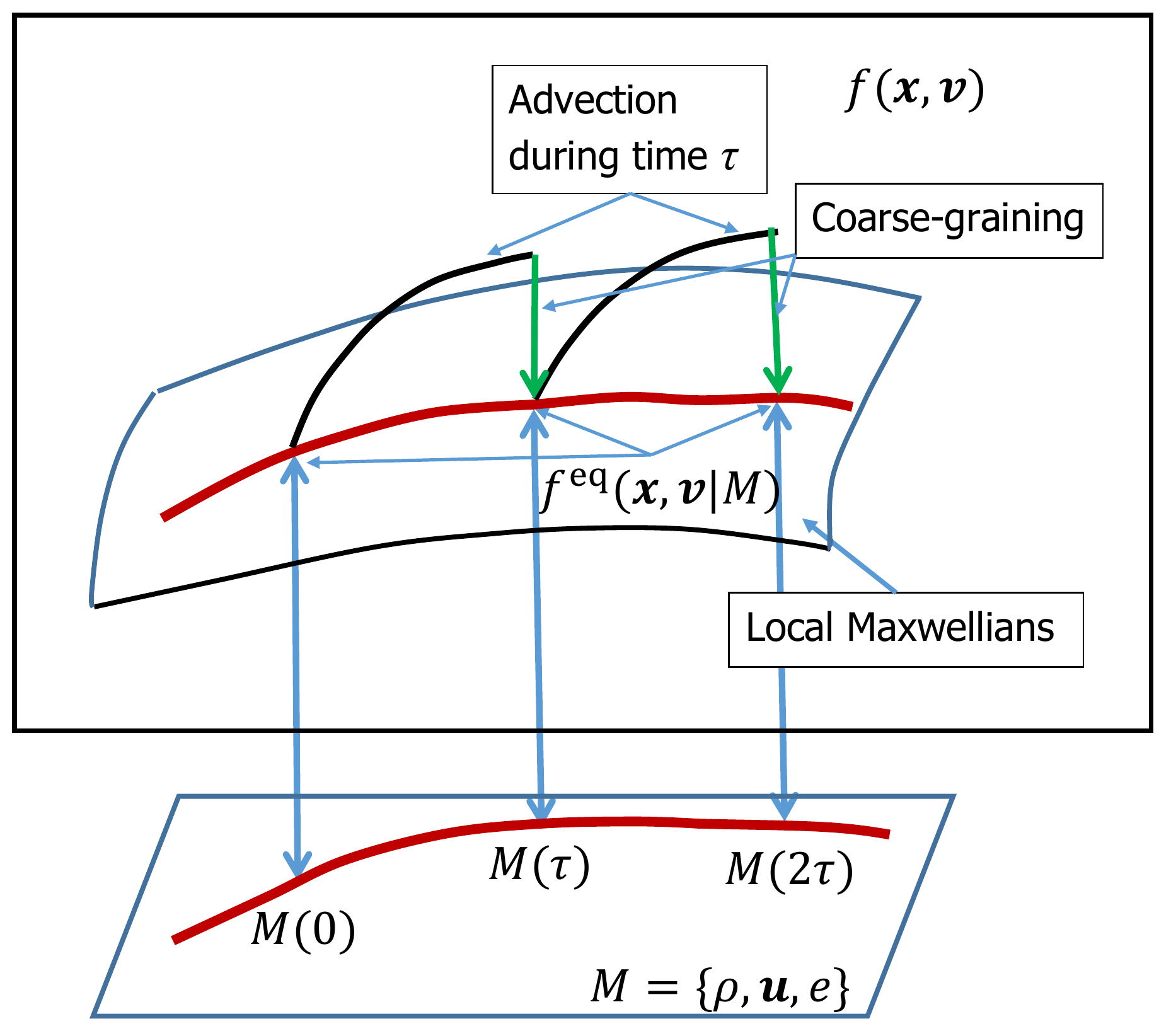}}
\caption {\label{Ehrenfestics}{\em Ehrenfests' chain}. Two spaces are schematically represented: the space of distribution functions (the upper rectangle) and the space of hydrodynamic fields (the bottom parallelogram). The vertical lines with bidirectional arrows illustrate coupling between the hydrodynamic fields and the corresponding local Maxwellians. The coarse-graining  is the projection of a distribution function onto the corresponding local Maxwellian with preservation of the hydrodynamic fields. }
\end{centering}
\end{figure}

We generalize Ehrenfests' idea of   coarse-graining by combining the genuine motion with the periodic partial equilibration. The result is {\em Ehrenfests' chain} (Figure~\ref{Ehrenfestics})  \cite{GorKarIlgOtt2001}. The general theory with the proof of the entropy production formula and various examples can be found elsewhere \cite{GorKarOttTat2001,KarTatGorOtt2003}. Here we just define Ehrenfests' chain for the collisionless gas and present the results of the coarse-graining.

First of all, for each value of hydrodynamic variables, the {\em equilibrium} distribution should be defined: $f^{\rm eq}(\boldsymbol{v}|\rho, \boldsymbol{u}, e)$. It is the Maxwellian distribution. In the units used in this section (we count neither moles nor grams but particles) the Maxwellian is
\begin{equation}\label{LocalMaxw}
f^{\rm eq}(\boldsymbol{v}|\rho, \boldsymbol{u}, e)=\rho\left(\frac{3}{4 \pi e}\right)^{3/2}\exp\left(-\frac{3(v-u)^2}{4e}\right).
\end{equation}
For the given hydrodynamic fields,  $\rho(\boldsymbol{x}), \boldsymbol{u}(\boldsymbol{x}), e(\boldsymbol{x})$, we use the {\em local Maxwellian}, $ f^{\rm eq}(\boldsymbol{v}|\rho(\boldsymbol{x}), \boldsymbol{u}(\boldsymbol{x}), e(\boldsymbol{x}))$, that is a distribution in six-dimensional space of positions and velocities, which is the Maxwellian for each $\boldsymbol{x}$.

Define Ehrenfests' chain with time step $\tau$. Let the initial values of the hydrodynamic fields be given, $\rho(\boldsymbol{x},0), \boldsymbol{u}(\boldsymbol{x},0), e(\boldsymbol{x},0)$. The initial distribution is
$$f(\boldsymbol{x},\boldsymbol{v},0)= f^{\rm eq}(\boldsymbol{v}|\rho(\boldsymbol{x},0), \boldsymbol{u}(\boldsymbol{x},0), e(\boldsymbol{x},0)).$$
Start the advection from this distribution. After time $\tau$ it approaches the distribution $f(\boldsymbol{x}-\boldsymbol{v}\tau,\boldsymbol{v},0)$. The hydrodynamic fields $\rho(\boldsymbol{x},\tau), \boldsymbol{u}(\boldsymbol{x},\tau), e(\boldsymbol{x},\tau)$ are defined using the moments of this distribution: $\rho(\boldsymbol{x},\tau)=\int f(\boldsymbol{x}-\boldsymbol{v}\tau,\boldsymbol{v},0)  \, \D ^3 v$, etc.

If the hydrodynamic fields $\rho(\boldsymbol{x},k \tau), \boldsymbol{u}(\boldsymbol{x},k \tau), e(\boldsymbol{x},k \tau)$ are given then $$f(\boldsymbol{x},\boldsymbol{v},k \tau)= f^{\rm eq}(\boldsymbol{v}|\rho(\boldsymbol{x},k \tau), \boldsymbol{u}(\boldsymbol{x},0), e(\boldsymbol{x},k \tau)),$$
and the hydrodynamic fields  $\rho(\boldsymbol{x},(k+1) \tau), \boldsymbol{u}(\boldsymbol{x},(k+1) \tau), e(\boldsymbol{x},(k+1) \tau)$ are defined using the moments of the distribution $f(\boldsymbol{x}-\boldsymbol{v}\tau,\boldsymbol{v},k \tau)$. Ehrenfests' chain is the sequence $\rho(\boldsymbol{x},k \tau),\boldsymbol{u}(\boldsymbol{x},k \tau), e(\boldsymbol{x},k \tau)$ ($k=0,1,2,\ldots$) (Figure~\ref{Ehrenfestics}). The hydrodynamic fields do not change in the equilibration steps and their trajectories between the equilibration jumps are just `shadows' of the advection (\ref{freeflight}). Equilibration returns the distribution to the local Maxwellian form and free flight destroys this form.

As a result, a discrete-time dynamical system  is defined in the space of hydrodynamic fields. We can look for a continuous-time system which transforms  $\rho(\boldsymbol{x},k \tau),\, \boldsymbol{u}(\boldsymbol{x},k \tau),\, e(\boldsymbol{x},k \tau)$ into $\rho(\boldsymbol{x},(k+1) \tau),\, \boldsymbol{u}(\boldsymbol{x},(k+1) \tau), e(\boldsymbol{x},(k+1) \tau)$ during time $\tau$.

Let us describe formally this procedure. Denote five hydrodynamic fields as $M(t)=\{\rho(\boldsymbol{x},\,\boldsymbol{v},t),\, \boldsymbol{u}(\boldsymbol{x},\boldsymbol{v},t) ,\, e(\boldsymbol{x},\boldsymbol{v},t) $. We aim to find a closed system $\frac{\upartial}{\upartial t} M=J_{\tau}(M)$, where $J_{\tau}$ is a (nonlinear) operator, which depends on the coarse-graining time $\tau$ as a parameter, and $J_{\tau}(M)$ is also a set of five fields (scalar, vector, and scalar again). Let operator  $m$ map  the distribution $f$ onto hydrodynamic fields: $M=m(F)$ (\ref{moments}). Select an initial state: $M(0)=M_0$ and $f_0$ is the corresponding local Maxwellian, $f_0=f^{\rm eq}(\boldsymbol{x},\boldsymbol{v}|M_0)$. Represent the result of advection during time $\tau$ as a power series in $\tau$:
$$f(\tau)=f_0+\sum_{k=1}^{\infty} \frac{\tau^k}{k!} (-1)^k (\boldsymbol{v},\boldsymbol{\nabla})^k f_0.$$
Find the power series for the result of hydrodynamic motion according to a hypothetical equation $\dot{M}=J_{\tau}(M)$: $$M(t)=M(0)+tJ_{\tau}+\sum_{k=2}^{\infty}\frac{t^k}{k!}\left.\frac{\D^{k-1}J_{\tau}(M)}{\D t^{k-1}}\right|_{M=M_0},$$
where the sequence of derivatives  $\frac{\D ^{k-1}J_{\tau}(M)}{\D t^{k-1}}$ is calculated by iterations of the chain rule from the equation $\dot{M}=J_{\tau}(M)$. We use this series for $t=\tau$.

\begin{figure}
\begin{centering}
\boxed{\includegraphics[width=0.43\textwidth]{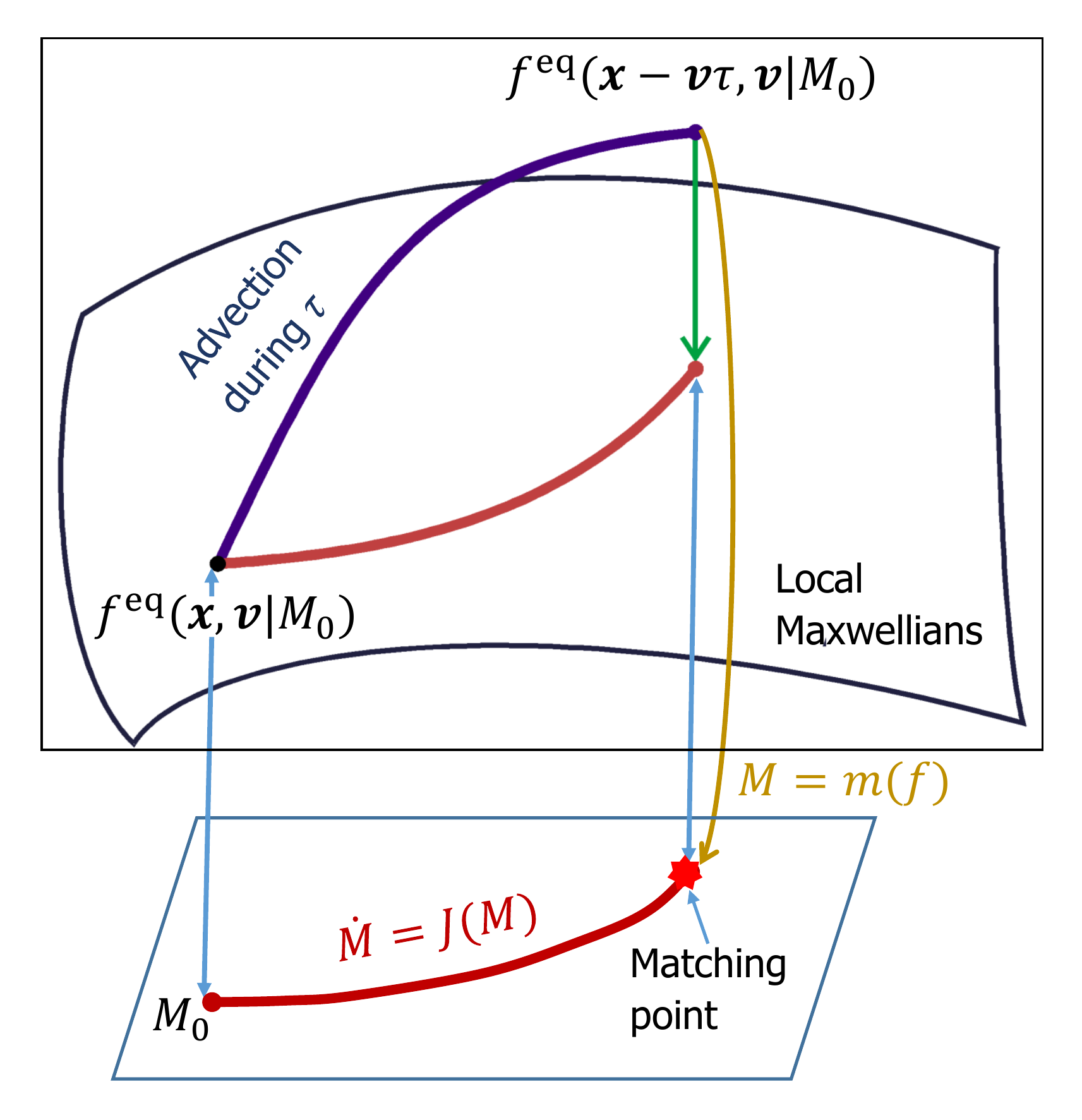}}
\caption {\label{EhrenfMatch}{\em Matching condition for hydrodynamic equations}. Two spaces are schematically represented: the space of distribution functions (upper rectangle) with the manifold of local Maxwellians and the space of hydrodynamic fields (the bottom parallelogram). The vertical lines with bidirectional arrows illustrate coupling between the hydrodynamic fields and the corresponding local Maxwellians. The matching condition is: $M(\tau)=f^{\rm eq}(\boldsymbol{x}-\boldsymbol{v}\tau, \boldsymbol{v}|M_0)$.}
\end{centering}
\end{figure}

Recall that $J_{\tau}(M)$ depends also on $\tau$ as on a parameter. Represent this dependence as a power series as well and substitute it into the series for $M(t)$ $(t=\tau)$. The matching condition should hold: $M(\tau)=m(f(\tau))$ (Fig.~\ref{EhrenfMatch}).
Coefficients at the same powers of $\tau$ in $M(\tau)$ and $m(f(\tau))$ should coincide.
As a result, we recover the macroscopic vector field  $J_{\tau}$ order-by-order.

The zeroth-order terms give the Euler equations and the first--order terms give the Navier--Stokes--Fourier equations with viscosity and heat conductivity  proportional to $\tau \rho e$ with constant coefficients ($\rho e$ is the density of internal energy): $\lambda=-\frac{2}{3}\mu$, $\mu=\frac{1}{3}\tau \rho e$ \cite{KarTatGorOtt2003}. These formulas recall the Maxwell mean free path estimations of kinetic coefficients: $\mu\approx \frac{1}{3}\rho l \bar{v}=\frac{1}{3}\rho \tau_{f} \bar{v}^2$  (where $\bar{v}$ is the thermal velocity and $\tau_{f}$ is the mean free time, see, for example \cite{Reif1965}).
The thermal velocity for Maxwellian distribution is  $\bar{v}\sim \sqrt{4e/3}$ (that is the most probable velocity, whereas  the  average magnitude of the velocity of the particles is $\frac{2}{\sqrt{\pi}} \bar{v}$). Therefore, the coarse-graining time $\tau$ can be considered as the analogue of the mean free time $\tau_{f}$ with a constant coefficient of order 1.

 It is worth mentioning that viscosity and thermal conductivity grow with the coarse-graining time, and dissipative processes become more intensive with the growth of the mean free path. Collisions in gas delay diffusive transport and decrease kinetic coefficients.

Thus, the simple model of non-equilibrium gas based on Ehrenfests' coarse graining produces the Navier--Stokes--Fourier equations in the first order of the coarse-graining time $\tau$. This model has only one free parameter,   $\tau$, and, therefore, the viscosity and the thermal conductivity are proportional (and Prandtl number Pr=1). This is not a miracle, and just demonstrates that the Navier--Stokes--Fourier equations provide an unavoidable first dissipative correction to the Euler equations. The problem is in the next approximation. In the second order of $\tau$, the coarse graining produces the Korteweg terms from the collisionless gas.

Consider the linearized fluid dynamics near a state with $\boldsymbol{u}=0$, $\rho=\rho_0$, $e=e_0$. Introduce the dimensionless deviation from this state: $$M_0=\frac{ \rho-\rho_0}{\rho_0}, \, M_i= \frac{u_i}{\bar{v}} \; (i=1,2,3), \, M_4=\frac{3}{2}\left(\frac{ \rho-\rho_0}{\rho_0} +\frac{ e-e_0}{e_0}\right),$$
where $\bar{v}=\sqrt{\frac{4}{3}e_0}$. Use the new space scale in which $\bar{v}=1$: $x_{\rm new}=x/\bar{v}$. In these variables, the linearized second approximation in $\tau$ for the coarse-grained free flight advection is:
\begin{equation}\label{EhrenfKortew}
\begin{split}
&\frac{\upartial M_0}{\upartial t}=\sum_{i=1}^3 \frac{\upartial M_i}{\upartial x_i};\\
&\frac{\upartial M_i}{\upartial t}=-\frac{1}{3}\frac{\upartial M_4}{\upartial x_i}+
 \frac{\tau}{4}\sum_{j=1}^3\frac{\upartial}{\upartial x_j}\left(\frac{\upartial M_i}{\upartial x_j}+
 \frac{\upartial M_j}{\upartial x_i} - \frac{2}{3}\delta_{ij}\sum_{k=1}^3  \frac{\upartial M_k}{\upartial x_k}\right)\\
 &\qquad +\tau^2 \frac{\upartial}{\upartial x_i}\left( \frac{1}{8}\Delta M_0 + \frac{89}{108}\Delta M_4     \right)      \; (i=1,2,3);\\
&\frac{\upartial M_4}{\upartial t}=-\frac{5}{2}\sum_{k=1}^3  \frac{\upartial M_k}{\upartial x_k}+ \frac{5\tau}{2}\sum_{k=1}^3 \Delta M_4+ \tau^2\frac{59}{9} \Delta\left( \sum_{k=1}^3 \frac{\upartial M_k}{\upartial x_k}\right).
\end{split}
\end{equation}
Here, $\Delta=\sum_{k=1}^3 \frac{\upartial^2 }{\upartial x_k^2}$ is the Laplace operator. Terms without $\tau$ correspond to the linearized Euler equation, the  first order terms give the Navier--Stokes--Fourier dissipation (viscosity and thermal conductivity). Terms of the second order in $\tau$ correspond to Korteweg's stress tensor (in equations for $M_i$, $i=1,2,3$) and to the contribution of capillarity into heat flux (in equation for $M_4$).

The parameter $\tau$  can be eliminated from equations (\ref{EhrenfKortew}) by rescaling. Recall that in these equations space and time are measured by the same time units (to provide $\bar{v}=1$). Let us select the new time and space unit $\tau$. In this scale, $\tau =1$  in (\ref{EhrenfKortew}). From some point of view, this means that the equations are the same for all values $\tau>0$. For example, it is sufficient to analyse stability just for one value $\tau=1$.

\begin{figure}
\begin{centering}
\boxed{a)\includegraphics[height=0.4 \textwidth]{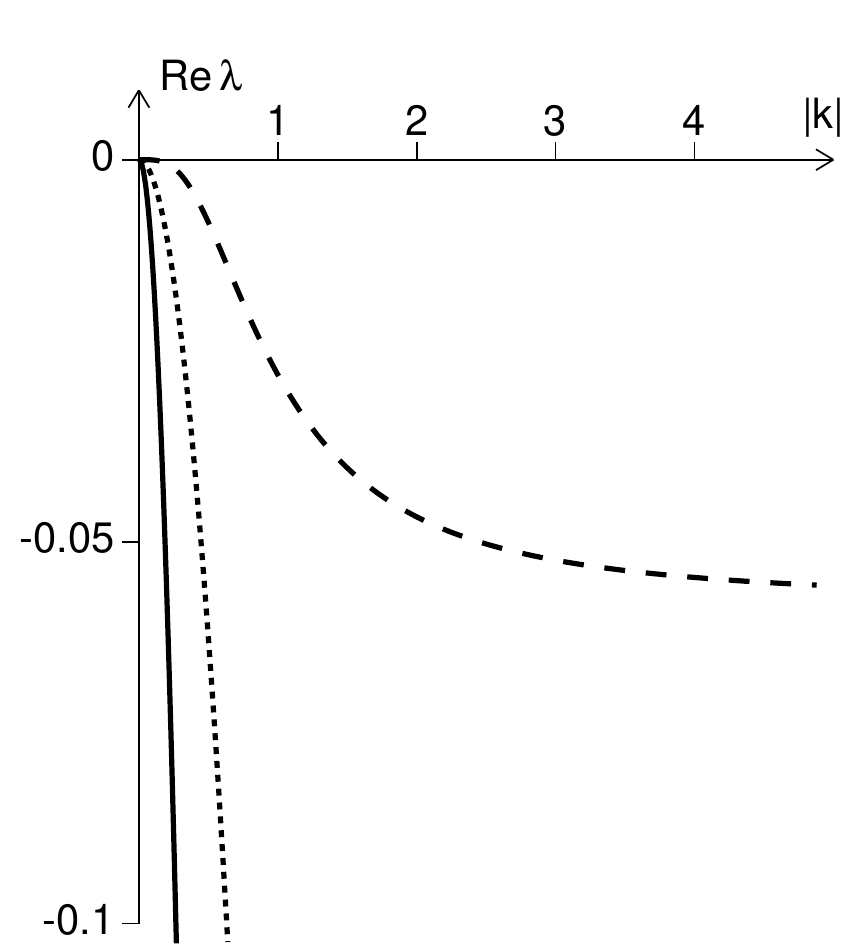}} \boxed{b)\includegraphics[height=0.4 \textwidth]{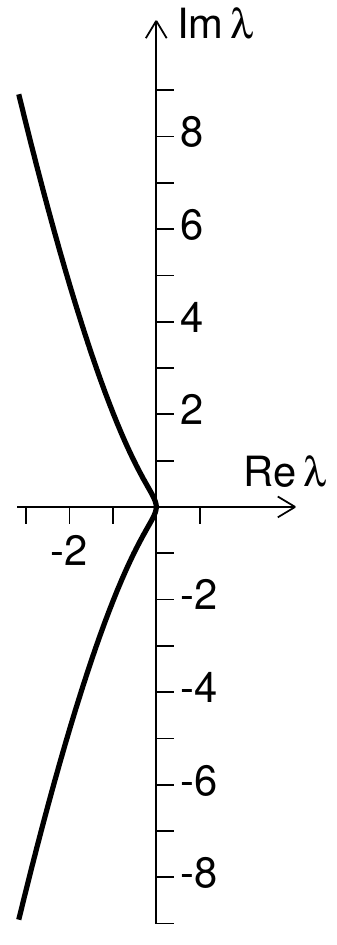}}
\caption {\label{SpectraKortewEhren}{\em Dispersion curves for the second approximation (\ref{EhrenfKortew})}. a) Dependence of attenuation rates on $|k|$:   solid $\lambda_{1,2}$, dashed $\lambda_{3}$, dotted $\lambda_{4,5}$; b) Curves $\lambda(k)$ on complex plane for $\lambda_{4,5}$}
\end{centering}
\end{figure}

Let us look for the solutions of  equations (\ref{EhrenfKortew}) in the form $M_j=A_j \exp(\lambda t+i \boldsymbol{kx})$.  Here, $\boldsymbol{k}$ is the wave vector and real parts of $\lambda$ describe dissipation. It is necessary for stability  that ${\rm Re} \lambda\leq 0$ for all real vectors $\boldsymbol{k}$. The characteristic equation for $\lambda(k)$
has five roots (Fig.~\ref{SpectraKortewEhren}). For all of them ${\rm Re} \lambda < 0$ ($k\neq 0$).
 In the short wave asymptotic of $\lambda$  ($k^2\to \infty $) the roots are:
\begin{equation}\label{EhrenRoots}
\frac{\lambda_{1,2}}{k^2}=-\frac{1}{4}; \; \frac{\lambda_{3}}{k^2}=O(1/k^2)\to 0;\;
\frac{\lambda_{4,5}}{k^2}=-\frac{17}{12}\pm i|k|\sqrt{\frac{59\cdot 89}{9\cdot 108}}.
\end{equation}

The second approximation in $\tau$ for the coarse-grained collisionless gas provides a sort of   `mean free path model' of capillarity effects and Korteweg stress in ideal gas. It seems to be surprising that such a simple approach generates the Korteweg equations (with some restrictions on the relationships between parameters, like Pr=1).

Nevertheless, two question remain: (i) are there capillarity effects in more realistic models of non-equilibrium gas, and (ii) what will happen in the coarse-grained collisionless gas beyond the second order in $\tau$? Of course, if we summarize the whole power series then the solutions of the resulting continuum mechanics equations will go exactly through the points of Erenfests' chain but what will happen  on the way?

The more realistic model of non-equilibrium gases is well-known, that is the Boltzmann equation.

\section{From Boltzmann kinetics to fluid dynamics: model reduction}

\subsection{The model reduction problem}

Let us consider {\em kinetic equations} which describe the evolution of a one-particle gas distribution function $f(\boldsymbol{x}, \boldsymbol{v},t)$
\begin{equation}\label{BOL0}\boxed{
    \frac{\upartial f}{\upartial t}+ \sum_i v_i \frac{\upartial f}{\upartial x_i}= \frac{1}{\rm Kn}Q(f).}
\end{equation}
The only difference from the free flight advection (\ref{advection}) is  the collision operator $Q(f)$ in the right hand part of (\ref{BOL0}). For the {\em Boltzmann equation}, $Q$ is a quadratic
operator and, therefore, the notation $Q(f,f)$ is often used. Kn is the {\em Knudsen number}, which  is a dimensionless parameter defined as $\mathrm {Kn}=\frac {l }{L}$, where $l$ is the mean free path and $L$ is `representative physical length' scale. It aims to measure how important  the {\em microscopic} effects are (associated with $l$) at the {\em macroscopic} scale ($L$).

The collision term $Q(f)$ in  (\ref{BOL0})  describes the change in distribution $f$ due to collisions. The term $Q(f)$ does not contribute directly into the time derivatives of the hydrodynamic variables, $\rho=\int f  \, \D ^3 v$, $\boldsymbol{u}=
\int \boldsymbol{v} f  \, \D ^3 v$ and $e= \frac{1}{2}\int (\boldsymbol{v-u})^2 f  \, \D ^3 v$ because, due to the mass, momentum and energy conservation in collisions:
$$\int \{1; \boldsymbol{v}; \boldsymbol{v}^2 \} Q(f)  \, \D ^3 v =0.$$
For the space-uniform distributions the collision operator in  (\ref{BOL0}) provides relaxation to Maxwellian distribution (\ref{LocalMaxw})  \cite{Carleman1933} (about the general problem of existence and stability for the Boltzmann equations we refer readers to \cite{DiPerna1989}).

\begin{figure}
\begin{centering}
\boxed{\includegraphics[width=0.7 \textwidth]{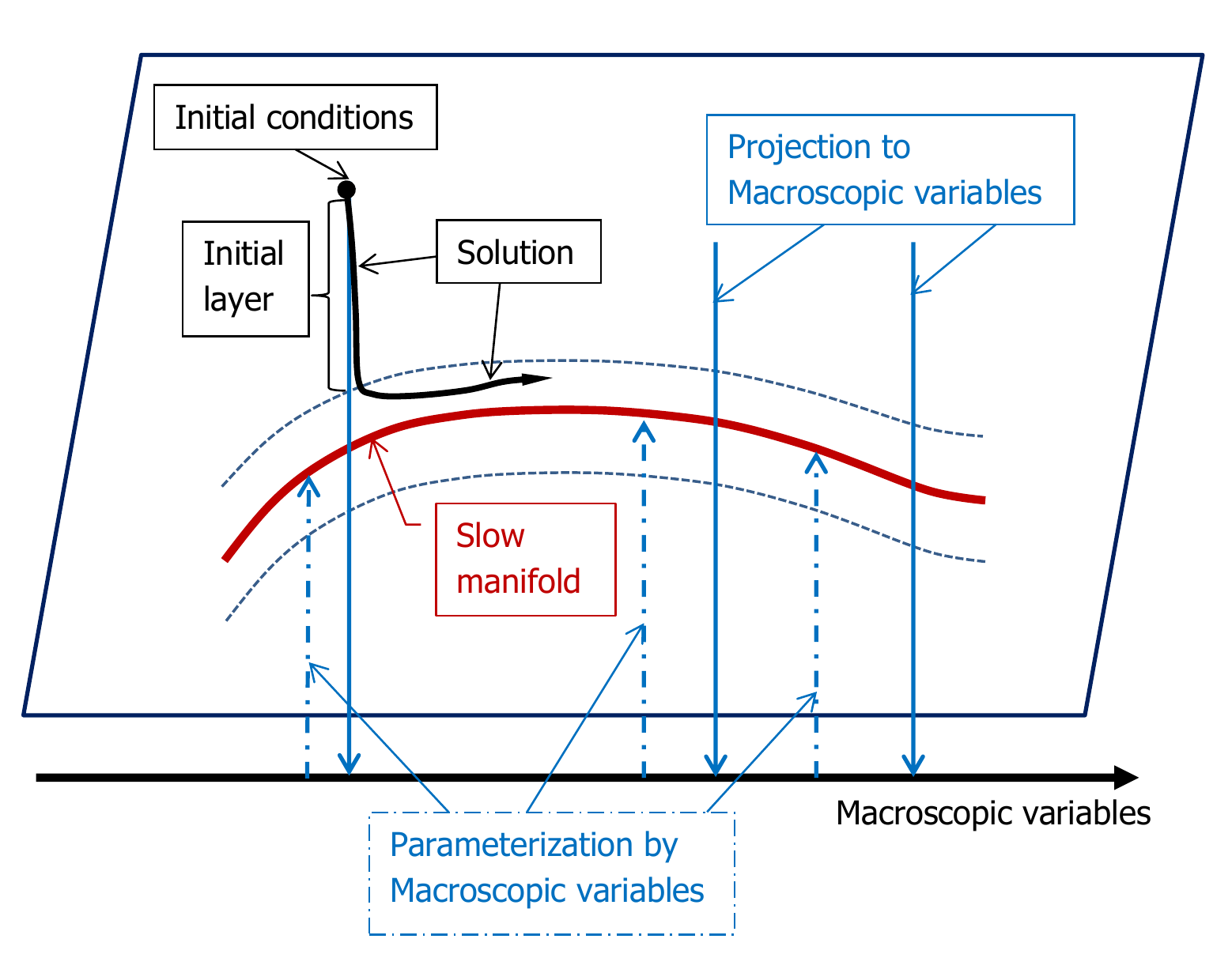}}
\caption {\label{fig1SLw} Fast--slow decomposition. Bold dashed lines outline the
vicinity of the slow manifold where the solutions stay after initial layer.
The projection of the distributions onto the hydrodynamic fields and the
parametrization of this manifold by the hydrodynamic fields are represented.  (A derivative work based on Fig.~1 from \cite{GorbanKarlin2014}.)}
\end{centering}
\end{figure}

Therefore, the following qualitative `nonrigorous picture of the Boltzmann dynamics' \cite{DescvilVill2005}  is expected for the solutions: (i) the
collision term goes quickly almost to its equilibrium (the system almost approaches a
local equilibrium) and during this fast initial motion the changes of hydrodynamic
variables are small, (ii) after that the distribution function is defined with high
accuracy by the hydrodynamic variables (if they have bounded space derivatives). The
relaxation of the collision term almost to its equilibrium is supported by monotonic
entropy growth (Boltzmann's $H$-theorem). This qualitative picture is illustrated in
Fig.~\ref{fig1SLw}.

\begin{figure}
\begin{centering}
\boxed{\includegraphics[width=0.7 \textwidth]{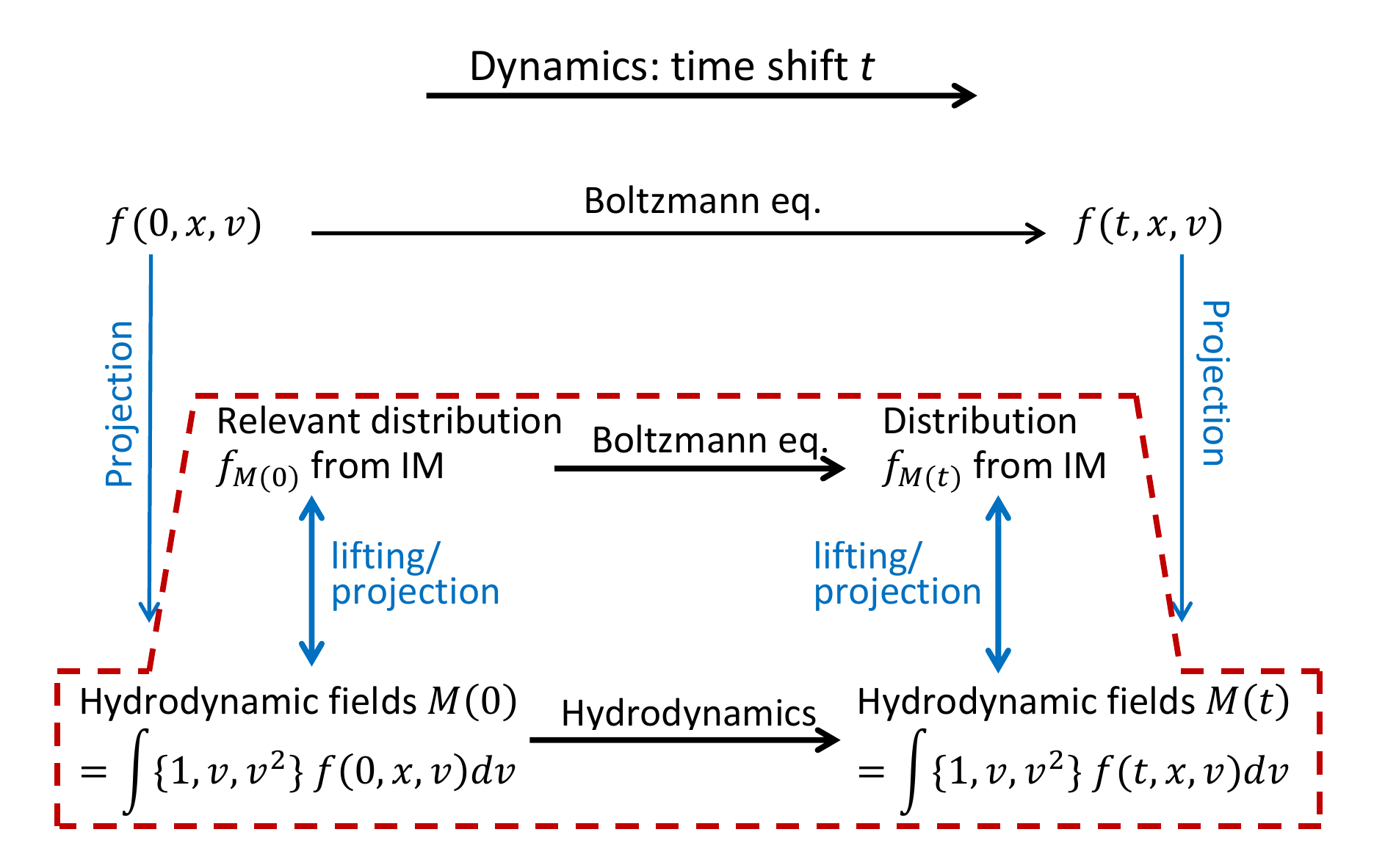}}
\caption {\label{McKeanDiag}{\em McKean diagram.} All the model reduction approaches aim to
create a {\em  lifting operation}, from the hydrodynamic variables to
the relevant distributions on the invariant manifold. IM stands for Invariant Manifold.
The part of the diagram in the dashed polygon is commutative, that is any superposition of operations following the arrows does not depend on the path but only on the start and end points. (A derivative work based on Fig.~2 from \cite{GorbanKarlin2014}.)}
\end{centering}
\end{figure}

Perhaps, McKean gave the first clear explanation of the problem as a construction of a
`nice submanifold' where `the hydrodynamical equations define the same flow as the (more
complicated) Boltzmann equation does' \cite{McKean1965}. He presented the problem by a
 diagram and we reproduce his idea in slightly revised form in
Fig.~\ref{McKeanDiag}. How to find this `nice submanifold'? Perhaps, the first task is to write an equation for them (for more detail we refer to the chapter `Invariance equation' of the book \cite{GorKarLNP2005}).

\subsection{Invariance equation \label{Sec:invariance}}

The invariance equation expresses the fact that the vector field is tangent to the
manifold. In 1892, A.M. Lyapunov introduced and studied this equation in his doctoral thesis `The general problem of the stability of motion' (the Lyapunov `auxiliary theorem' \cite{Lya1992}, which is recently used in various applications, from model reduction in  chemical kinetics to  optimal control problems \cite{KazKra1998,KazGood2002}).

The essence of the invariance equation and the Chapman--Enskog method become more transparent in the abstract form when the bulky details of the genuine Boltzmann equation do not obscure the simple geometric sense. Let us consider an equation
in a domain $U$ of a normed space $E$ with analytical
right hand side
\begin{equation}\label{AbstractEq}
\upartial_t f = J(f).
\end{equation}
We call $f$ a {\em microscopic variable}. A space of {\em macroscopic variables} (moment fields) is defined with a linear map  $m: f \mapsto M$ ($M$ are macroscopic variables). Assume that the image of $m$ is the whole space of $M$ (i.e. $m$ is surjective). We are looking for an invariant manifold (McKean's nice manifold') parameterized with the macroscopic fields $M$. For such manifolds we use the notation $\boldsymbol{f}_M$. The {\em self-consistency condition} $m(\boldsymbol{f}_{M})=M$ is necessary: the manifold is parametrized by its own value of macroscopic variables.

For the reduction of Boltzmann's kinetics hydrodynamics, the microscopic variable is the one-particle distribution function $f$ and the macroscopic variables are hydrodynamic fields. In extended irreversible thermodynamics (EIT) the larger sets of macroscopic variables are considered \cite{EIT}.

The invariance equation for $\boldsymbol{f}_M$ is
\begin{equation}\label{IM}
\boxed
{J(\ff_M)=(D_M \ff_M) m(J(\ff_M)).}
\end{equation}
Here, the differential $D_M$ of $\ff_M$ is calculated at the point $M=m(\ff_M)$.

Equation (\ref{IM}) means that the time derivative of $\ff$ on the manifold $\ff_M$ can be calculated by a simple chain rule:
calculate the derivative of $M$ using the map $m$, $\dot{M}=m(J(\ff_M))$,  then write that on the invariant manifold the time dependence  $\ff(t)$
can be expressed through the time dependence of $M(t)$: $\ff(t)=\ff_{M(t)}$ and the time derivatives also coincide.
If we find the approximate solution to eq. (\ref{IM}) then the approximate reduced model (hydrodynamics) is
\begin{equation}\label{projectedMIM}
\upartial_t M=m(J(\ff_M)) .
\end{equation}
The invariance equation can be represented in the form
\begin{equation}\label{Micro/Makro}
\upartial^{\rm micro}_t \ff_M = \upartial^{\rm macro}_t \ff_M,
\end{equation}
where the microscopic time derivative, $\upartial^{\rm micro}_t \ff_M$ is just a value of
the vector field $J(\ff_M)$ and the macroscopic time derivative is calculated by the
chain rule, $$\upartial^{\rm macro}_t \ff_M= (D_M \ff_M) \upartial_t M $$ under the
assumption that dynamics of $M$ follows the projected  equation (\ref{projectedMIM}).
For more details we refer to the review paper \cite{GorbanKarlin2014}.

\subsection{Chapman--Enskog expansion in brief \label{Sec:Ch-E}}

The Chapman--Enskog approach assumes the special
singularly perturbed structure of the equations and looks for the invariant manifold in a
form of the series in the powers of a small parameter $\epsilon$. A one-parametric system
of equations is considered:
\begin{equation}\label{AbstractEqSingPert}
\upartial_t f +A(f)=\frac{1}{\epsilon}Q(f).
\end{equation}
The following assumptions connect the macroscopic variables to the singular perturbation:
\begin{itemize}
\item{$m(Q(f))=0$;}
\item{for each $M\in m(U)$ the system of equations
$Q(f)=0$, $m(f)=M$
has a unique solution $\ff^{\rm eq}_M$ (in Boltzmann kinetics it is the local
Maxwellian);}
\item{$\ff^{\rm eq}_M$ is asymptotically stable and globally attracting for the fast system
$\upartial_t f =\frac{1}{\epsilon}Q(f)$
in $(\ff^{\rm eq}_M+\ker m)\cap U$. }
\end{itemize}
Let the $\mathcal{Q}_M$ be the differential of the fast vector field $Q(f)$ at equilibrium $\ff^{\rm eq}_M$: $\mathcal{Q}_M=(D_MQ(\ff))_{\ff=\ff^{\rm eq}_M}$. For the Chapman--Enskog method it is important that $\mathcal{Q}_M$ is invertible in $\ker m$.

The invariance equation for the singularly perturbed system (\ref{AbstractEqSingPert}) with the moment parametrization $m$
is:
\begin{equation}\label{AbstractEqSingIM}
\boxed{
\frac{1}{\epsilon}Q(\ff_M)= A(\ff_M)-(D_M \ff_M) (m(A(\ff_M))).
}
\end{equation}
 The self-consistency condition $m(\ff_M)=M$ gives $m(D_M
\ff_M)m(J)=m(J)$ for all $J$, hence,
\begin{equation}\label{self-consistency}
m[A(\ff_M)-(D_M \ff_M) m(A(\ff_M))]=0.
\end{equation}
If we find an approximate solution $\ff_M$ of (\ref{AbstractEqSingIM}) then the corresponding
macroscopic (hydrodynamic) equation (\ref{projectedMIM}) is

\begin{equation}\label{projectedMIMSing}
\upartial_t M+m(A(\ff_M))=0 .
\end{equation}

Let us represent all the operators in (\ref{AbstractEqSingIM}) by the Taylor series (for
the Boltzmann equation $A$ is the linear free flight operator, $A=v\cdot \nabla$,    and
$Q$ is the quadratic collision operator).  We look for the invariant manifold in the form
of the power series:
\begin{equation}\label{ChapmanEnskogGeneral}
\ff_M=\ff^{\rm eq}_M +\sum_{i=1}^{\infty} \epsilon^i \ff^{(i)}_M
\end{equation}
with the self-consistency condition $m(\ff_M)=M$, which implies $m(\ff^{\rm eq}_M)=M$,
$m(\ff^{(i)}_M )=0$ for $i\geq 1$. After matching the coefficients of the series in
(\ref{AbstractEqSingIM}), we obtain for every $\ff^{(i)}_M$ a linear equation
\begin{equation}\label{ChEeq}
\mathcal{Q}_M \ff^{(i)}_M = P^{(i)}(\ff^{\rm eq}_M,\ff^{(1)}_M, \ldots , \ff^{(i-1)}_M),
\end{equation}
where the polynomial operator $ P^{(i)}$ at each order $i$ can be obtained by
straightforward calculations from (\ref{AbstractEqSingIM}). Due to the self-consistency,
$m(P^{(i)})=0$ for all $i$ and the equation (\ref{ChEeq}) is solvable. The first term of
the Chapman--Enskog expansion has a simple form
\begin{equation}\label{ChE1}
\boxed{
\ff^{(1)}_M=\mathcal{Q}_M^{-1} (1-(D_M \ff^{\rm eq}_M)m) (A(\ff^{\rm eq}_M)) .
}
\end{equation}
Many books and papers are devoted to the detailed analysis of this formula for the Boltzmann equation and
other kinetic equations after the classical book \cite{Chapman}. Most of the physical applications of kinetic theory, from transport processes in gases to modern numerical methods (lattice Boltzmann models
\cite{Succi2001}) give examples of the practical applications and deciphering of this formula.
For the Boltzmann kinetics, the zero-order approximation, $\ff^{(0)}_M\approx\ff^{\rm
eq}_M$ produces  in projection on the hydrodynamic fields (\ref{projectedMIMSing}) the
compressible Euler equation. The first-order approximate invariant manifold,
$\ff^{(1)}_M\approx\ff^{\rm eq}_M+\epsilon \ff^{(1)}_M$,  gives  the compressible
Navier-Stokes equation and provides the explicit dependence of the transport coefficients
on the collision model.

The calculation of higher order terms needs nothing but differentiation and calculation of the inverse operator $\mathcal{Q}_M^{-1}$, although it may be rather bulky.  The second
order in $\epsilon$ hydrodynamic equations (\ref{projectedMIM}) are called Burnett
equations (with $\epsilon^2$ terms) and super-Burnett equations for higher orders.

Alas, the Burnett equations produce non-physical effects, instability of short waves (Bobylev's instability) and negative viscosity at the space scale near mean free path, i.e. close to the scales, where they are needed. What will happen if we sum up the whole Chapman--Enskog series (at least, hypothetically)? For Ehrenfests' chain the sum is expected to coincide with the chain in discrete time moments $n\tau$. For the Boltzmann equation, it remains unclear. Let us choose a simplified system, for which the model reduction can be performed explicitly for all time scales.

\subsection{Exact analytic solution of reduction problem for a simple kinetic equation \label{Sec:exact}}

The simplest model and the starting point in our analysis is
\begin{equation}
\label{Grad101}\boxed{
\begin{split}
\upartial_t p &=-\frac{5}{3}\upartial_x u,\\
\upartial_t u &=-\upartial_x p -\upartial_x \sigma,\\
\upartial_t \sigma &=-\frac{4}{3}\upartial_x u
-\frac{1}{\epsilon}\sigma ,
\end{split}}
\end{equation}

To obtain this system from Boltzmann's equation we have to select the set of macroscopic variables:
``Hydrodynamic fields plus stress tensor'. For these  10 variables (1 -- density, plus 3 -- momentum density, plus 1 -- energy, and plus 5 -- traceless symmetric stress tensor) find the representative (quasiequilibrium) density function $f$ by conditional maximization of entropy, $S=-\int f \ln f  \, \D ^3 v $, for given values of  these 10 macroscopic moments. The  conditional maximization of entropy subject to given first and second moments is a standard exercise in the MaxEnt approach. The result is the Gaussian distribution (more precisely, it is the locally Gaussian distribution  with space-dependent parameters defined by hydrodynamic fields).   Substitute this function into Botzmann's equation, calculate the corresponding time derivatives of macroscopic variables and the closed 10-moment MaxEnt Grad system is ready.
Linearize and study the solutions that depend on one space coordinate $x$ with the
velocity  oriented along the $x$ axis. Use dimensionless variables. (Here, $\sigma$ is the dimensionless $xx$-component of the stress tensor.)

Let us illustrate the invariance equation and the Chapman--Enskog series on the simplest example (\ref{Grad101}).
\begin{eqnarray*}
\ff=\left(
\begin{array}{c} p(x)\\u(x)\\ \sigma(x) \end{array}
\right), \;
m=\left(
\begin{array}{ccc} 1 & 0 & 0 \\ 0 & 1 & 0 \end{array}
\right), \;
M=\left(
\begin{array}{c} p(x)\\u(x) \end{array}
\right), \;
\ker m=\left\{\left(\begin{array}{c} 0\\ 0\\ y \end{array}
\right)\right\},
\end{eqnarray*}
\begin{eqnarray*}
A(\ff)=\left(
\begin{array}{c} \frac{5}{3}\upartial_x u \\ \upartial_x p +\upartial_x \sigma \\ \frac{4}{3}\upartial_x u \end{array}
\right),\;
Q(\ff)=\left(
\begin{array}{c} 0\\ 0\\ -\sigma \end{array}
\right),  \;
\mathcal{Q}_M^{-1}=\mathcal{Q}_M=-1 \mbox{ on } \ker m, \\
\end{eqnarray*}
\begin{eqnarray*}
\ff^{\rm eq}_M=\left(
\begin{array}{c} p(x)\\u(x)\\ 0 \end{array}
\right), \;
D_M \ff^{\rm eq}_M=
\left(
\begin{array}{cc} 1 & 0  \\ 0 & 1  \\ 0 & 0  \end{array}
\right), \;
\ff^{(1)}_M=\left(
\begin{array}{c} 0 \\ 0 \\ -\frac{4}{3}\upartial_x u \end{array}
\right).\;
\end{eqnarray*}
We hasten to remark that (\ref{Grad101}) is a simple linear system and can be integrated
immediately in explicit form. However, that solution contains both the fast and slow
components and it does not readily reveal the slow hydrodynamic manifold of the system.
Instead, we are interested in extracting this slow manifold by a direct method. The
Chapman-Enskog expansion is thus the for this which we shall address
first.

The projected equations in the zeroth (Euler) and the first (Navier--Stokes) order  of
$\epsilon$ are
\begin{equation*}
\label{Grad101Eu}
\mbox{ (Euler) } \begin{array}{ll}\upartial_t p = -\frac{5}{3}\upartial_x u,\\
\upartial_t u =-\upartial_x p;
\end{array} \;\;\;
\mbox{ (Navier-Stokes) } \begin{array}{ll}
\upartial_t p =-\frac{5}{3}\upartial_x u,\\
\upartial_t u =-\upartial_x p +\epsilon \frac{4}{3}\upartial_x^2 u.
\end{array}
\end{equation*}
It is straightforward to calculate the two next terms (Burnett and super-Burnett ones) but
let us introduce convenient notations to represent the whole Chapman-Enskog series for
(\ref{Grad101}). Only the third component of the invariance equation
(\ref{AbstractEqSingIM}) for (\ref{Grad101}) is non-trivial because of the self-consistency
condition (\ref{self-consistency}), and we can write
\begin{equation}\label{Drad101IM}
-\frac{1}{\epsilon}\sigma_{(p,u)}= \frac{4}{3}\upartial_x u-\frac{5}{3}(D_p \sigma_{(p,u)})(\upartial_x u)-(D_u \sigma_{(p,u)}) (\upartial_x p +\upartial_x \sigma_{(p,u)}).
\end{equation}
Here, $M={(p,u)}$ and the differentials are calculated by the elementary rule: if a
function $\Phi$ depends on values of $p(x)$ and its derivatives, $\Phi=\Phi(p,\upartial_x
p,\upartial^2_x p,\ldots)$   then  $D_p \Phi$ is a differential operator,
$$D_p \Phi = \frac{\upartial \Phi}{\upartial p}+\frac{\upartial \Phi}{\upartial (\upartial_x p)}\upartial_x+\frac{\upartial \Phi}{\upartial (\upartial_x^2 p)}\upartial^2_x+\ldots$$

The equilibrium of the fast system (the Euler approximation) is known,
$\sigma_{(p,u)}^{(0)}=0$. We have already found
$\sigma_{(p,u)}^{(1)}=-\frac{4}{3}\upartial_x u$ (the Navier--Stokes approximation). In
each order of the Chapman--Enskog expansion $i\geq 1$ we  get from (\ref{Drad101IM}):
\begin{equation}\label{Grad101CE}
\sigma_{(p,u)}^{(i+1)}= \frac{5}{3}(D_p \sigma_{(p,u)}^{(i)})(\upartial_x u)+(D_u \sigma_{(p,u)}^{(i)})(\upartial_x p)
+ \sum_{j+l=i} (D_u \sigma_{(p,u)}^{(j)})(\upartial_x \sigma_{(p,u)}^{(l)}).
\end{equation}

This chain of equations is nonlinear but every $\sigma_{(p,u)}^{(i+1)}$ is a linear
function of derivatives of $u$ and $p$ with constant coefficients because this sequence
starts from $ -\frac{4}{3}\upartial_x u $  and the induction step in $i$ is obvious.

\begin{figure}
\begin{centering}
\boxed{a)\includegraphics[width=0.75 \textwidth]{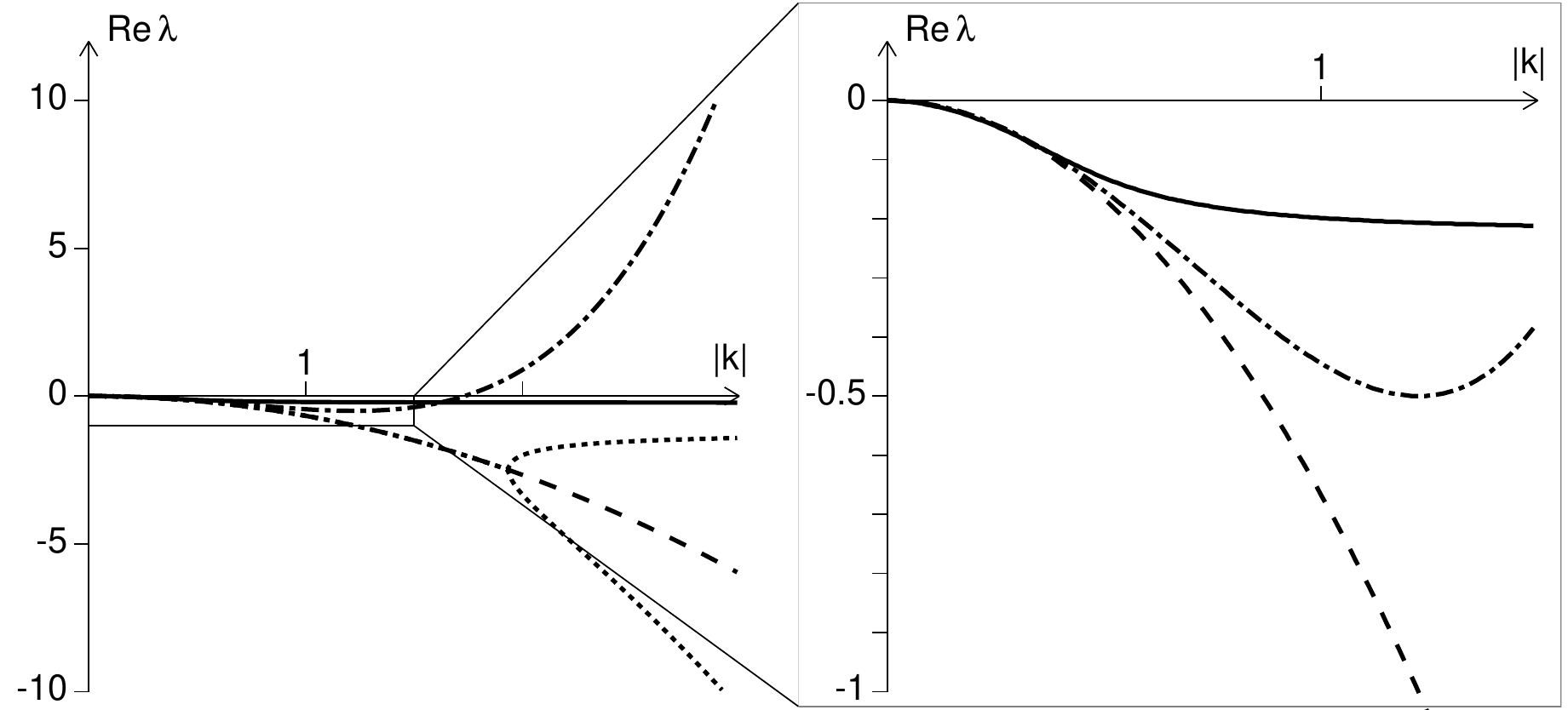}}\\ \vspace{3mm}\boxed{b)\includegraphics[width=0.75 \textwidth]{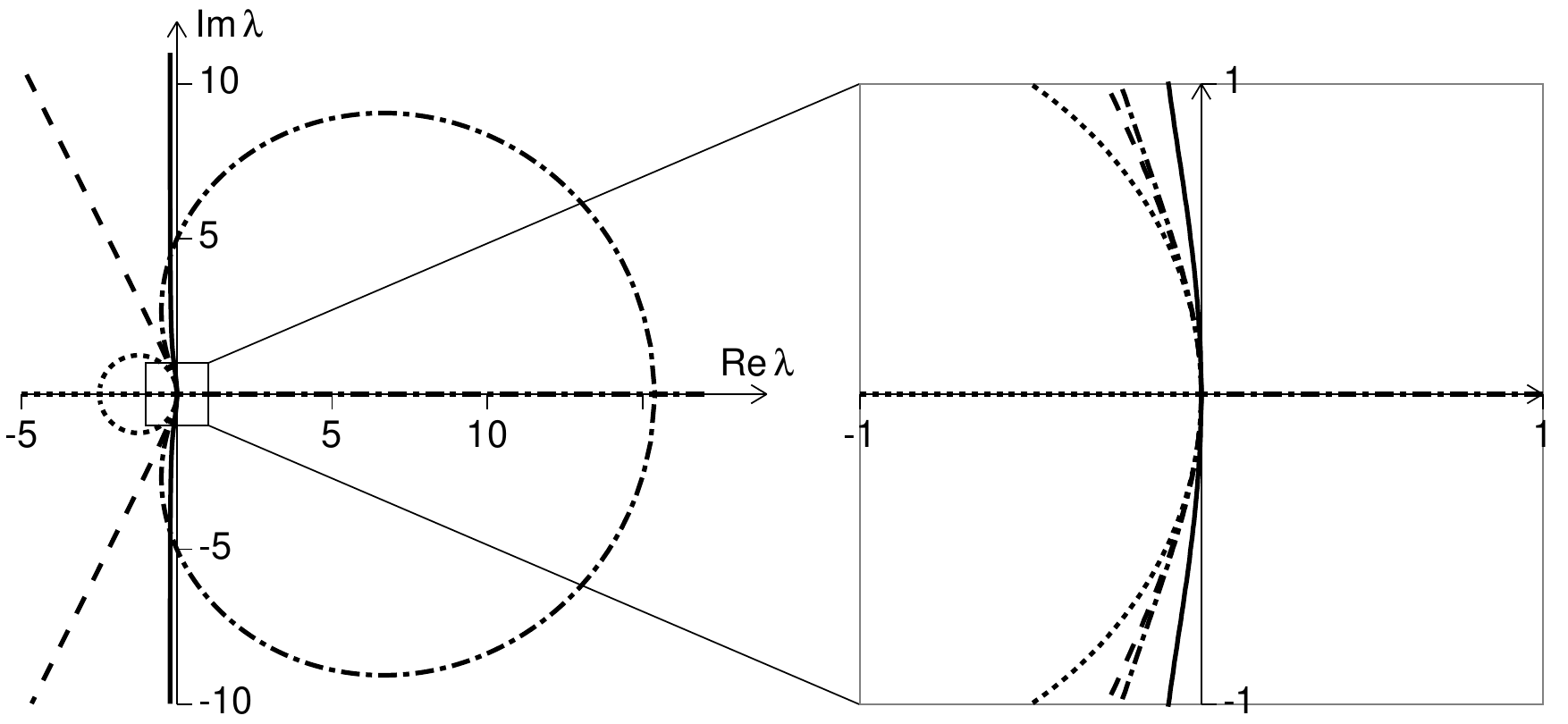}}
\caption {\label{SpectraSimplest}Dispersion curves (\ref{dispersionns101}) for various hydrodynamic approximations obtained from the simple kinetic equation (\ref{Grad101}):   solid - exact solution,  dotted -- Navier--Stokes approximation, dashed - Burnett equation, dash and dotted -- super Burnett  approximation.  a) Dependence of attenuation rates on $|k|$  (for the Navier--Stokes and Burnett curves ${\rm Re} \lambda$ coincide if $4k^2<15$; they differ in ${\rm Im} \lambda$); b) Curves $\lambda(k)$ on complex plane.}
\end{centering}
\end{figure}

Simple algebra gives for the Burnett term ($i+1=2$)
$\sigma_{(p,u)}^{(2)}=-\frac{4}{3}\upartial^2_x p $ and for the super Burnett term  ($i+1=3$) the
 $\sigma_{(p,u)}^{(3)}= -\frac{4}{9}\upartial^3_x u$.
The projected equations have the form
\begin{eqnarray}
\label{Grad101Burn}
&\begin{array}{ll}\upartial_t p = -\frac{5}{3}\upartial_x u,\\
\upartial_t u =-\upartial_x p +\epsilon \frac{4}{3}\upartial_x^2 u + \epsilon^2 \frac{4}{3}\upartial_x^3 p;
\end{array} \mbox{ Burnett; } \\
&\begin{array}{ll}
\upartial_t p =-\frac{5}{3}\upartial_x u,\\
\upartial_t u =-\upartial_x p +\epsilon \frac{4}{3}\upartial_x^2 u + \epsilon^2 \frac{4}{3}\upartial_x^3 p + \epsilon^3 \frac{4}{9}\upartial_x^4 u;
\end{array} \mbox{ super Burnett. }\label{Grad101supBurn}
\end{eqnarray}
Compute the dispersion relation for
these hydrodynamic modes. Exclude $\epsilon$ using a new space-time scale, $x^{\prime}=\epsilon^{-1} x$, and
$t^{\prime}=\epsilon^{-1} t$. Look for wave solutions $u=u_k \varphi(x^{\prime},t^{\prime})$,
and $p=p_k \varphi(x^{\prime},t^{\prime})$, where
$\varphi(x^{\prime},t^{\prime})=\exp(\lambda t^{\prime}+ikx^{\prime})$, and $k$ is a
real-valued wave vector. The following dispersion relations $\lambda(k)$ are
the conditions of a non-trivial solvability of the corresponding linear system with
respect to $u_k$ and $p_k $:
\begin{equation}\label{dispersionns101}
\begin{split}
\lambda_{\pm}&=-\frac{2}{3}k^2 \pm \frac{1}{3}i|k|\sqrt{15-4k^2}, \;\; \mbox{Navier--Stokes}; \\
\lambda_{\pm}&=-\frac{2}{3}k^2 \pm \frac{1}{3}i|k|\sqrt{15+16 k^2}, \;\; \mbox{Burnett};\\
\lambda_{\pm}&=-\frac{2}{9}k^2 (3-k^2) \pm \frac{1}{9}i|k|\sqrt{135+144k^2+24k^4- 4k^6 },\;\;  \mbox{super Burnett};
\end{split}
\end{equation}
For the super Burnet equations ${\rm Re} \lambda>0$  if $k^2>3$. This is an example of Bobylev's instability. The dispersion curves are presented in Fig.~\ref{SpectraSimplest}.

Let us  analyze the structure of the Chapman--Enskog series given by
the recurrence formula (\ref{Grad101CE}). The terms in the series alternate: For odd $i=1,3,\ldots$ they are
proportional to $\upartial_x^i u$ and for even $i=2,4,\ldots$ they are proportional to
$\upartial_x^i p$. It follows from the parity properties: $u$ and $\upartial_x$ change sign after spatial reflection (vectors), whereas $p$ (a scalar) and $\sigma$ (a second-order tensor) are invariant with respect to inversion. This global structure of the Chapman--Enskog series gives the following representation of the stress
$\sigma$ on the hydrodynamic invariant manifold
\begin{equation}\label{Structure1D10M}
\sigma(x)=A(-\upartial_x^2)\upartial_x u(x) +B(-\upartial_x^2) \upartial_x^2 p(x),
\end{equation}
where $A(y)$, $B(y)$ are yet unknown functions and the sign `$-$' in the arguments is
adopted for simplicity of formulas in the Fourier transform.
For the stress tensor (\ref{Structure1D10M}) the reduced  equations have the form
\begin{equation}
\label{Grad101reduced}\boxed{
\begin{split}
\upartial_t p &=-\frac{5}{3}\upartial_x u,\\
\upartial_t u &=-\upartial_x p -\upartial_x [A(-\upartial_x^2)\upartial_x u + B(-\upartial_x^2) \upartial_x^2 p].
\end{split}}
\end{equation}

\subsection{Exact invariant manifold in Fourier representation\label{IEFourier}}

It is convenient to work with the pseudodifferential operators like
(\ref{Structure1D10M}) in Fourier space. Let us denote $p_k$, $u_k$ and
$\sigma_k$, where $k$ is the `wave vector' (space frequency).

The Fourier-transformed kinetic equation (\ref{Grad101}) takes the form
($\epsilon=1$):
\begin{equation}\label{Grad101F}
\begin{split}
\upartial_t p_k &=-\frac{5}{3}ik u_k ,\\
\upartial_t u_k &=-ik p_k -ik \sigma_k ,\\
\upartial_t \sigma_k &=-\frac{4}{3}ik u_k
-\sigma_k .
\end{split}
\end{equation}

We know already that the result of the reduction should be a function $\sigma_k (u_k ,
p_k ,k)$ of the following form:
\begin{equation}\label{Parametriz1D10M}
\sigma_{k}(u_k , p_k ,k) =ikA(k^2)u_k -k^2 B(k^2)p_k ,
\end{equation}
where $A$ and $B$ are unknown real-valued functions of $k^2$.

The solution of the invariance equation amounts to finding the two
functions, $A(k^2 )$ and $B(k^2 )$. Let us rewrite the invariance equation for these unknown
functions. Compute the time derivative of $\sigma_{k}(u_k , p_k ,k)$
in { two } different ways. First, use the right hand side of the third equation in
(\ref{Grad101F}). We find the {\em microscopic} time derivative:
\begin{equation}
\label{101microderivat}
 \upartial_t^{\rm micro}\sigma_k =-ik\left(\frac{4}{3}+A\right)u_k
+k^2 Bp_k .
\end{equation}

Secondly, let us use the chain rule
and the first two equations in (\ref{Grad101F}). We find
 the {\em macroscopic} time derivative:
\begin{equation}
\begin{split}\label{101macroderivat}
\upartial_t^{\rm macro} \sigma_k & =\frac{\upartial
\sigma_k }{\upartial u_k }\upartial_t u_k +\frac{\upartial \sigma_k
}{\upartial p_k }\upartial_t p_k \\  &=ikA\left(-ikp_k
-ik\sigma_k \right)-k^2 B\left(-\frac{5}{3}iku_k \right)\\
&=ik\left(\frac{5}{3}k^2 B+k^2 A\right)u_k +k^2 \left(A-k^2
B\right)p_k .
\end{split}
\end{equation}

The microscopic time derivative should coincide with the
macroscopic time derivative for all values of $u_k$ and $p_k$.
This is the invariance equation:
\begin{equation}\label{FourierInvEq1D10M}\boxed{
\upartial_t^{\rm macro} \sigma_k =\upartial_t^{\rm
micro}\sigma_k.}
\end{equation}
 For the kinetic system (\ref{Grad101F}), it reduces to a system of two
quadratic equations for functions $A(k^2)$ and $B(k^2)$:
\begin{equation}\label{AlgebraicIM1D10M}
-A-\frac{4}{3}-k^2 \left(\frac{5}{3}B+A^2\right)=0,\; \;  -B+A\left(1-k^2 B\right)=0.
\end{equation}
Solving the system (\ref{AlgebraicIM1D10M}) for $B$, and introducing a new function,  $X(k^2 )=k^2 B(k^2 )$, we obtain an equivalent cubic equation:
\begin{equation}
\label{factor101}
-\frac{5}{3}(X-1)^2 \left(X+\frac{4}{5}\right)=\frac{X}{k^2 }.
\end{equation}
We need  the real-valued functions $A(k^2)$ and $B(k^2)$ (\ref{Parametriz1D10M}).
The real-valued root $X(k^2 )$ of (\ref{factor101}) is unique and { negative} for all finite values $k^2 $. Moreover, the function $X(k^2 )$ is a monotonic function of $k^2 $ and
\begin{equation}
\label{limits101}
\lim_{|k|\rightarrow0}X(k^2 )=0, \quad
\lim_{|k|\rightarrow\infty}X(k^2 )=-0.8.
\end{equation}
Therefore, both $B=k^2X$ and $A=B/(1-X)$ are negative.

For the given $A(k^2)$ and $B(k^2)$ the dispersion relation
for the hydrodynamic modes are (just use the stress tensor $\sigma$ (\ref{Parametriz1D10M}) in the first two equations of system (\ref{Grad101F}) and express $A$ and $B$ through $X$) :
\begin{equation}
\label{dispersion101}
\lambda_{\pm}=
\frac{X}{2(1-X)}\pm i\frac{|k|}{2}\sqrt{\frac{5X^2
-16X +20}{3}},
\end{equation}
where $X=X(k^2 )$ is the real-valued root of equation (\ref{factor101}). Since $0 > X(k^2
)>-1$ for all $|k|>0$, the attenuation rate, ${\rm Re}( \lambda_{\pm})$, is negative for
all $|k|>0$, and the exact acoustic spectrum of the reduced equations {\it is
stable for arbitrary wave lengths} (Fig.~\ref{SpectraSimplest}, solid line). In the
short-wave limit, from (\ref{limits101}), (\ref{dispersion101}) we obtain the {\em saturation of dissipation}:
\begin{equation}
\label{limit101}
\lim_{|k|\rightarrow\infty}{\rm Re} \lambda_{\pm}=-\frac{2}{9} ;\ \;\; \lim_{|k|\rightarrow\infty} \frac{{\rm Im} \lambda_{\pm}}{|k|} = \pm
\sqrt{3}.
\end{equation}

Thus, we found the invariant hydrodynamic manifold in two steps:
\begin{enumerate}
\item{We used the invariance equation, Chapman--Enskog procedure and the
    symmetry properties to find a linear space where the hydrodynamic invariant
    manifold is located. This space is parameterized by two functions of one
    variable (\ref{Parametriz1D10M});}
\item{We used the invariance equation again  and defined an algebraic manifold in this
    space. For the simple kinetic system (\ref{Grad101}), (\ref{Grad101F}) this
    manifold is given by the system of two quadratic equations which depend linearly
    on $k^2$ (\ref{AlgebraicIM1D10M}).}
\end{enumerate}

\section{Van der Waals capillarity energy in ideal gas \label{Sec:CapIdGas}}

\subsection{The energy formula and `capillarity' of ideal gas\label{Sec:caplillarity}}

Let us look on the stress tensor (\ref{Parametriz1D10M}). Traditionally, $\sigma$ in kinetics of gases is considered as a viscous stress tensor but the second term,
$B(-\upartial_x^2) \upartial_x^2 p(x)$, is proportional to second derivative of $p(x)$ and
it does not meet usual expectations ($\sigma \sim \nabla u$). Slemrod
\cite{SlemQuaterly2012,SlemCAMWA2013} noticed that the proper interpretation of  this
term is the capillarity tension rather than the viscous stress. This is made clear by inspection
of the energy equation. Let us derive  the energy equation for the simple
model (\ref{Grad101}). Find the time derivative of the kinetic energy due to the first two
equations (\ref{Grad101}):
\begin{equation}
\begin{split}
\frac{1}{2}  \upartial_t \int_{-\infty}^{\infty} u^2 \,\D x &= \int_{-\infty}^{\infty} u \upartial_t u \, \D x
=-\int_{-\infty}^{\infty} u\upartial_x p \, \D x- \int_{-\infty}^{\infty} u\upartial_x \sigma \, \D x \\ &=
- \frac{1}{2} \upartial_t \frac{3}{5} \int_{-\infty}^{\infty} p^2 \, \D x + \int_{-\infty}^{\infty}  \sigma \upartial_x u\, \D x .
\end{split}
\end{equation}
Here we used integration by parts under the standard  assumptions at infinitiy. Note, that $\frac{1}{2}\upartial_t (p^2)=-\frac{5}{3}p\upartial_x u$.

In $x$-space the energy equation has the standard form (\ref{energy}):
\begin{equation}\label{energy101x}
\frac{1}{2} \upartial_t \left(\frac{3}{5}\int_{-\infty}^{\infty} p^2 \, \D x+\int_{-\infty}^{\infty} u^2 \,\D x\right)=
\int_{-\infty}^{\infty} \sigma  \upartial_x  u\, \D x.
\end{equation}
Note that the usual factor $\rho$ in front of
$u^2$ is absent because we work with linearized equations and dimensionless variables.

Let us use in (\ref{energy101x}) the representation (\ref{Structure1D10M}) for $\sigma$
and notice that $\upartial_x   u= -\frac{3}{5}\upartial_t p$:
$$\int_{-\infty}^{\infty} \sigma  \upartial_x   u\, \D x
=\int_{-\infty}^{\infty} (\upartial_x u)(A(-\upartial_x^2)\upartial_x u) \, \D x
-\frac{3}{5}\int_{-\infty}^{\infty} (\upartial_t p) [ B(-\upartial_x^2) \upartial_x^2 p ] \,
\D x. $$
 The operator $B(-\upartial_x^2) \upartial_x^2$ is symmetric, therefore,
$$\int_{-\infty}^{\infty} (\upartial_t p) [ B(-\upartial_x^2) \upartial_x^2 p ] \, \D
x=\frac{1}{2}\upartial_t\left(\int_{-\infty}^{\infty} p [ B(-\upartial_x^2) \upartial_x^2 p
] \, \D x\right). $$ The quadratic form,
\begin{equation}\label{CapillarEnergyX}
U_c=\frac{3}{5}\int_{-\infty}^{\infty} p ( B(-\upartial_x^2)
\upartial_x^2 p ) \, \D x= - \frac{3}{5}\int_{-\infty}^{\infty} (\upartial_x p )( B(-\upartial_x^2)
\upartial_x p )\, \D x
\end{equation}
may be considered as a part of the energy. Moreover, the function $B(y)$ is negative, hence,
this form is  positive.
Finally, the energy formula in $x$-space is
\begin{equation}\label{EnergyXfin}
\boxed{\begin{split}
 \frac{1}{2} \upartial_t \int_{-\infty}^{\infty} \left(\frac{3}{5} p^2+u^2 -
\frac{3}{5} (\upartial_x p )( B(-\upartial_x^2)
\upartial_x p ) \right) \, \D x \\
=\int_{-\infty}^{\infty} (\upartial_x u)(A(-\upartial_x^2)\upartial_x u) \, \D x.
\end{split}}
\end{equation}

It is crucially important that the functions $A(k^2)$ and $B(k^2)$ are negative, indeed,  despite of the fact that some of the Taylor coefficients may be positive and, therefore, the truncation of the formula at some powers of $\upartial_x$ may not work. We have to use either the whole series or special approximations which preserve negativity of $A$ and $B$.

Slemrod \cite{SlemQuaterly2012}  represents the structure of the obtained energy formula as
\begin{equation}\label{CapillIdea}\boxed{
\begin{split}
\upartial_t ({\rm MECHANICAL\ ENERGY})+\upartial_t
({\rm CAPILLARITY\ ENERGY})\\ = {\rm VISCOUS\
DISSIPATION}.
\end{split}}
\end{equation}

\subsection{Matched asymptotics: from $k^2=0$ to $k^2=\infty$ \label{HighFreqAs}}

For large values of $k^2$, an analogue of the Chapman--Enskog expansion at an
infinitely-distant point is useful. Let us rewrite the algebraic equation for the
invariant manifold (\ref{AlgebraicIM1D10M}) in the form
\begin{equation}\label{AlgebraicIM1D10M1/k^2}
\frac{5}{3}B+A^2=-\varsigma (\frac{4}{3}+A),\;\;
AB=\varsigma(A-B),
\end{equation}
where $\varsigma=1/k^2$. For the analytic solutions near the point $\varsigma=0$ the
Taylor series is: $A=\sum_{l=1}^{\infty} \alpha_l \varsigma^l$, $B=\sum_{l=1}^{\infty}
\beta_l \varsigma^l$, where $\alpha_1=-\frac{4}{9}$, $\beta_1=-\frac{4}{5}$,
$\alpha_2=\frac{80}{2187}$, $\beta_2=\frac{4}{27}$, ... . The first term gives for the
frequency (\ref{dispersion101}) the same limit:
\begin{equation}
\label{dispersion101infty}
\lambda_{\pm}=-\frac{2}{9} \pm i {|k|}{\sqrt{3}},
\end{equation}
and the higher-order term give some corrections.

Let us match the Navier--Stokes term and the first term in the $1/k^2$ expansion. Find rational functions $A\approx\tilde{A}(k^2)$ and $B\approx\tilde{B}(k^2)$ such that $\tilde{A}(0)=\tilde{B}(0)=-\frac{4}{3}$  (the Navier--Stokes limit) and $k^2 \tilde{A}(k^2) \to -\frac{4}{9}$,  $k^2 \tilde{B}(k^2) \to -\frac{4}{5}$ when $k^2 \to \infty$ (the short wave limit). Solution with the minimal powers of $k^2$ is:
\begin{equation}\label{MergedAsymptotics}
A\approx-\frac{4}{3+9 k^2}, \;\; B\approx-\frac{4}{3+5k^2}
\end{equation}
and
\begin{equation}
\sigma_{k}=ikA(k^2)u_k -k^2 B(k^2)p_k \approx -\frac{4ik}{3+9k^2}u_k +\frac{4k^2}{3+5k^2}p_k .
\end{equation}

This simplest non-locality captures the main effects: the Navier--Stokes approximation for  small Knudsen and Mach numbers (small $k^2$)  and the proper asymptotic for short waves
(large $k^2$) with the saturation of dissipation. This saturation is a universal effect \cite{Rosenau1989,KurganovRosenau1997,GorKarKhleb1991,GKTTSP94,SlemrodSatur1998} and hydrodynamics that  do not take it  into account cannot pretend to be an universal asymptotic
equation.

For the matched asymptotic (\ref{MergedAsymptotics}) we obtain from (\ref{Grad101reduced})
\begin{equation}
\label{Grad101reducedM}\boxed{
\begin{split}
&\upartial_t p =-\frac{5}{3}\upartial_x u,\\
&(1-3\upartial_x^2)\left(1-\frac{5}{3}\upartial_x^2\right)\upartial_t u =-\upartial_x p \\&\;\;\;\;+ \frac{4}{3}\upartial_x\left[\left(1-\frac{5}{3}\upartial_x^2\right)\upartial_x u + (1-3\upartial_x^2) \upartial_x^2 p\right],
\end{split}}
\end{equation}

These equations give us a clue about the proper asymptotic of the continuum mechanic equations for rarefied non-equilibrium gas: we can expect the appearance of several factors of the form $(1-\alpha \Delta)$, where $\Delta$ is the Laplace operator.

\section{Other approaches: conclusion and outlook}

We presented the main continuum mechanics equations for compressible fluids, from the Euler  to  the Navier--Stokes--Fourier and Korteweg equations. The problem of deduction of the proper equations for highly non-equilibrium fluxes was formulated. The essential part of Hilbert's sixth problem is model reduction from kinetics to continuum mechanics \cite{SlemCAMWA2013,GorbanKarlin2014}. We demonstrated two classical approaches: solution of the invariance equation (\ref{IM}),  (\ref{Micro/Makro}) by the Chapman--Enskog series and the Ehrenfests coarse graining.  We solved the reduction problem exactly for a simple kinetic system  (\ref{Grad101}) . This system provided us with a benchmark for comparison of various methods and for the explicit demonstration of the van der Waals capillarity of ideal gases (\ref{EnergyXfin}),  (\ref{CapillIdea}).

There are many attempts to solve the reduction problem and deduce the continuum mechanics equations for non-vanishing Knudsen number from the Boltzmann equation. We can solve the invariance equation by the direct Newton (or Newton--Kantorovich) method \cite{GKTTSP94}. The Newton iterations for the invariance equations provide much better results  than the Chapman--Enskog expansion. The first iteration gives the Navier--Stokes  asymptotic for long waves and the qualitatively correct behavior with saturation  for short waves. The second iteration gives the proper higher order approximation  in the long wave limit and the quantitatively proper asymptotic for short waves.

Another idea is extension of the set of independent variables. Grad proposed to write the equations for higher moments \cite{Grad1949}. His method in combination with the Maximum Entropy approach  got the name `Extended irreversible thermodynamics' (EIT) \cite{EIT}. We can start from any equation of EIT and apply the method of invariant manifold: write the invariance equation (\ref{IM}), find the first terms of the Chapman--Enskog expansion, etc. \cite{KGDNPRE98}. There appears also a group of methods, which take account of some of the higher order terms in the lower order truncation of the Chapman--Enskog expansion \cite{Bobylev2005,Sone2012,Struchtrup2005}. These terms may regularize the singularities in the lower orders.

A rich family of mesoscopic lattice Boltzmann methods was developed for applications in fluid dynamics and beyond \cite{Succi2001}. They can be successfully applied to microfluidics and various other problems between fluid dynamics and kinetics.

Most of these methods lead beyond continuum mechanics. The Korteweg equation is the first post Navier--Stokes--Fourier equation, which remains inside continuum mechanics but captures some nonequilibrium kinetic phenomena like the capillarity of ideal gas. Exact solutions of the reduction problem (from kinetics to hydrodynamics) give us hints of how the post Navier--Stokes equations may look.

Finally, it should be noted that the dissipation and the capillarity terms in the energy equation resulting from the exact summation are of same order if the gradients are not small. Thus,  Korteweg's term is not just a small correction to dissipation but rather a contribution to the energy balance on the same scale. Korteweg's equations were originally introduced in relation to non-ideal gas equation of state to capture the effect of surface tension between different phases.

So, after all, {\it why}  capillarity emerges in ideal gas? The answer to this question is in the nature of the interface of the {\em brick of matter} in Cauchy stress construction. Whenever the gradients of the hydrodynamic fields become commensurable with the mean free path, there is an energy price to be paid for their maintenance. The highly idealized conventional picture of continuous media assumes an almost impenetrable elastic interface (Euler) with only a small smearing (Navier--Stokes--Fourier) of the order of a mean free path. However, when the gradients increase, also the dispersion effects come into play which is precisely what the surface energy is responsible for in Korteweg's picture. It is clear that the non-locality associated with this effect is essential since no truncation of the Chapman--Enskog series is possible.

We demonstrated how the van der Waals capillarity appears in dynamics of ideal gas. The above assertion is based on the linearized model equations, the only case so far which was amenable to the exact solution. The nonlinear case still requires work but the useful clues are provided by Korteweg's equations and kinetic models with exactly solvable reduction problem.

\section*{Acknowledgements}

ANG was supported by the EPSRC grant EP/N022653/1, IVK acknowledges support by the ERC grant 291094-ELBM and SNSF grant 200021\_149881.

\section*{Notes on contributors}

\begin{wrapfigure}{l}{0.17\textwidth}
\includegraphics[width=0.15\textwidth]{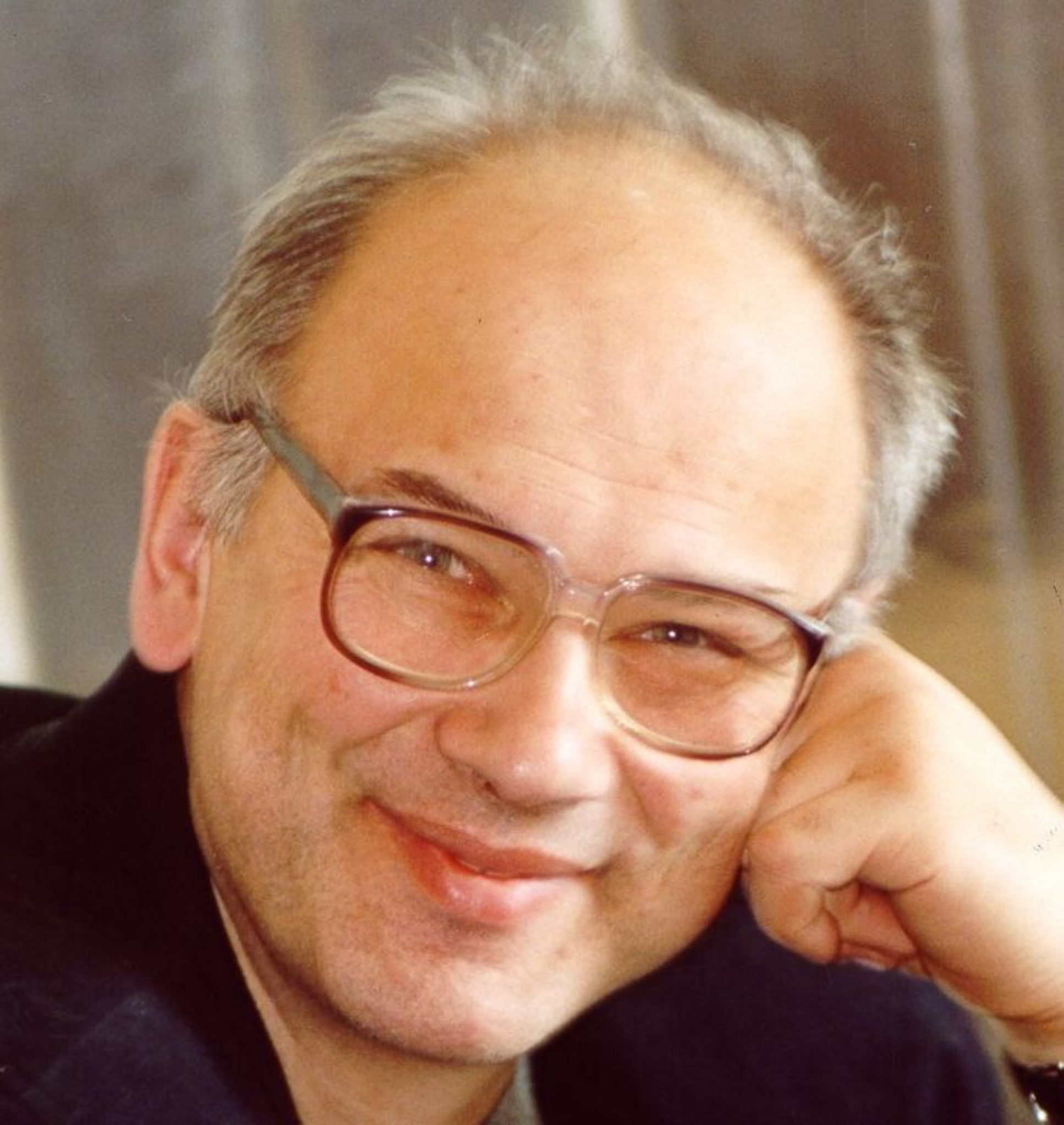}
\end{wrapfigure}Alexander N. Gorban holds a personal chair in Applied
Mathematics at the University of Leicester since 2004. He had worked
for Russian Academy of Sciences, Siberian Branch (Krasnoyarsk, Russia)
and ETH Z\"urich (Switzerland), had been a visiting professor and
research scholar at Clay Mathematics Institute (Cambridge, US), IHES
(Bures--sur-Yvette, \^Ile de France), Courant Institute of
Mathematical Sciences (NY, US) and Isaac Newton Institute for
Mathematical Sciences (Cambridge, UK). Main research interests:
Dynamics of systems of physical, Chemical and biological kinetics;
Biomathematics; Data mining and model reduction problems.

\vspace{4mm}
\begin{wrapfigure}{l}{0.17\textwidth}
\includegraphics[width=0.15\textwidth]{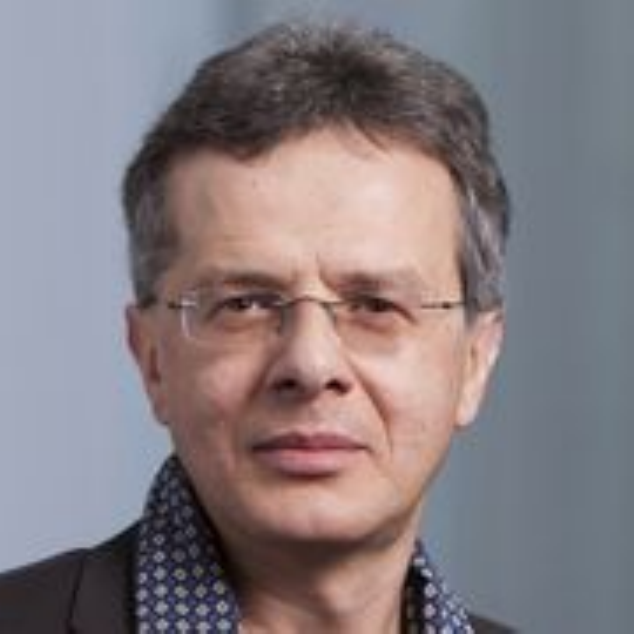}
\end{wrapfigure}Ilya Karlin is Faculty Member at the Department of Mechanical and
Process Engineering, ETH Zurich, Switzerland. He was Alexander von Humboldt
Fellow at the University of Ulm (Germany), CNR Fellow at the Institute of
Applied Mathematics CNR ``M. Picone'' (Rome, Italy), and Senior Lecturer
in Multiscale Modeling at the University of Southampton (England).
Main research interests: Exact and non-perturbative results in kinetic theory;
Fluid dynamics; Entropic lattice Boltzmann method;
Model reduction for combustion systems.

\end{document}